\documentclass[journal]{IEEEtran}

  \usepackage{amssymb}
  \usepackage{latexsym}
  \usepackage{amsfonts}
  \usepackage{amsthm}
  \usepackage{amsmath}
  \usepackage{graphics}
  \usepackage{setspace}
  \usepackage{graphicx}
  \usepackage{epstopdf}
  \usepackage{color}
  \usepackage{float}
  \usepackage{wrapfig}
  \usepackage{color}
  \usepackage{rotating}
  \usepackage{array} 
  \usepackage{dblfloatfix}
  \usepackage[hyphens]{url}
  \usepackage[table]{xcolor}
  \usepackage{multirow}

  \usepackage{graphics}
  \usepackage{setspace}
  \usepackage{algorithm}
  \usepackage{algorithmic}
  \usepackage{graphicx}
  \usepackage{epstopdf}  
  \usepackage{color}
  \usepackage{float}
  \usepackage{wrapfig}
  \usepackage{color}
  
  \usepackage{graphicx}
  \usepackage{caption}
  \usepackage{subcaption}
  \usepackage{setspace}
  
  \definecolor{blue}{rgb}{0,0,0}

\usepackage{lipsum}
  \usepackage{amsmath}

\definecolor{litered}{RGB}{246,194,197}
\definecolor{liteyellow}{RGB}{250,227,199}
\definecolor{litegreen}{RGB}{206,235,183}

\newcommand{\llr}{\cellcolor{litered}} 
\newcommand{\lly}{\cellcolor{liteyellow}} 
\newcommand{\llg}{\cellcolor{litegreen}}

\newcommand{\gc}{\cellcolor[gray]{0.8}}

\ifCLASSINFOpdf
\else
\fi
\begin{document}
\setstretch{0.99}
\title{Sensor Search Techniques for Sensing as a Service Architecture for The Internet of Things}

\author{\IEEEauthorblockN{Charith Perera~\IEEEmembership{Student Member,~IEEE}, Arkady Zaslavsky~\IEEEmembership{Member,~IEEE}, Chi Harold Liu~\IEEEmembership{Member,~IEEE},
\\Michael Compton,  Peter Christen and Dimitrios Georgakopoulos \IEEEmembership{Member,~IEEE}}

%\vspace{-25pt}

%\IEEEauthorblockA{\IEEEauthorrefmark{1}Research School of Computer Science, The Australian National University, Canberra, ACT 0200, Australia}
%\IEEEauthorblockA{\IEEEauthorrefmark{2}CSIRO ICT Center, Canberra, ACT 2601, Australia}
%\IEEEauthorblockA{\IEEEauthorrefmark{3}IBM Research - China, Beijing, China \vspace{-25pt}}

%\author{\IEEEauthorblockN{Charith Perera\IEEEauthorrefmark{1}\IEEEauthorrefmark{2},
% Arkady Zaslavsky\IEEEauthorrefmark{2},
% Chi Harold Liu\IEEEauthorrefmark{3},
%Michael Compton\IEEEauthorrefmark{2},
%\\ Peter Christen\IEEEauthorrefmark{1} and
%Dimitrios Georgakopoulos\IEEEauthorrefmark{2}\vspace{-25pt}}

\thanks{An earlier version of this paper was accepted for oral presentation at the  IEEE 14th International Conference on Mobile Data Management (MDM), June 3--6, 2013, Milan, Italy, and has been accepted for publication in its proceedings.} %and was accepted to be published in its Proceedings.}

\thanks{C.~Perera, A.~Zaslavsky, M. Compton and D~Georgakopoulos are with the Information and Communication Centre, Commonwealth Scientific and Industrial Research Organisation,  Canberra, ACT, 2601, Australia (e-mail: firstname.lastname@csiro.au)}% <-this % stops a space
\thanks{P.~Christen is with the Research School of Computer Science, The Australian National University, Canberra, ACT 0200, Australia. (e-mail: peter.christen@anu.edu.au)}% <-this % stops a space
\thanks{C. H. Liu is with IBM Research---China, Beijing, China. (e-mail: chiliu@cn.ibm.com)}% <-this % stops a space

\thanks{Manuscript received xxx xx, xxxx; revised xxx xx, xxxx.}}


\markboth{IEEE SENSORS JOURNAL, VOL. xx, NO. xx, xxxxxxx xxxx}%
{Shell \MakeLowercase{\textit{et al.}}: Bare Demo of IEEEtran.cls for Journals}



\maketitle

\begin{abstract}
%\boldmath

\textcolor{blue}{The Internet of Things (IoT) is part of the Internet of the future and will comprise billions of intelligent communicating ``things'' or Internet Connected Objects (ICO) which will have sensing, actuating, and data processing capabilities. Each ICO will have one or more embedded sensors that will capture potentially enormous amounts of data. The sensors and related data streams can be clustered physically or virtually, which raises the challenge of searching and selecting the right sensors for a query in an efficient and effective way. This paper proposes a context-aware sensor search, selection and ranking model, called CASSARAM, to address the challenge of efficiently selecting a subset of relevant sensors out of a large set of sensors with similar functionality and capabilities. CASSARAM takes into account user preferences and considers a broad range of sensor characteristics, such as reliability, accuracy, location, battery life, and many more. The paper highlights the importance of sensor search, selection and ranking for the IoT, identifies important characteristics of both sensors and data capture processes, and discusses how semantic and quantitative reasoning can be combined together. This work also addresses challenges such as efficient distributed sensor search and relational-expression based filtering. CASSARAM testing and performance evaluation results are presented and discussed.}

\end{abstract}


% \fi
\IEEEpeerreviewmaketitle

%\vspace{-5pt}

\section{Introduction}
\label{sec:Introduction}
\IEEEPARstart{T}{he} number of sensors deployed around the world is increasing at a rapid pace. These sensors continuously generate enormous amounts of data. However, collecting data from all the available sensors does not create additional value unless they are capable of providing valuable insights that will ultimately help to address the challenges we face every day (e.g. environmental pollution management and traffic congestion management). Furthermore, it is also not feasible due to its large scale, resource limitations, and cost factors. When a large number of sensors are available from which to choose, it becomes a challenge and a time consuming task to select the \textit{appropriate}\footnote{We describe the term \textit{appropriate} in Section \ref{sec:Problem_Definition_and_Motivation}.} sensors that will help the users to solve their own problems.

\textcolor{black}{The sensing as a service \cite{ZMP003}} model is expected to be built on top of the IoT infrastructure and services. It also envisions that sensors will be available to be used over the Internet either for free or by paying a fee through midddleware solutions. Currently, several middleware solutions that are expected to facilitate such a model are under development. OpenIoT \cite{P377}, GSN \cite{P022}, and xively (xively.com) are some examples. These middleware solutions strongly focus on connecting sensor devices to software systems and related functionalities \cite{P377}. However, when more and more sensors get connected to the Internet, the search functionality becomes critical.

This paper addresses the problem mentioned above as we observe the lack of focus on sensor selection and search in existing IoT solutions and research. Traditional web search approach will not work in the IoT sensor selection and search domain, as text based search approaches cannot capture the critical characteristics of a sensor accurately. Another approach that can be followed is that of metadata annotation. Even if we maintain metadata on the sensors (e.g. stored in a sensor's storage) or in the cloud, interoperability will be a significant issue. Furthermore, a user study done by Broring et al. \cite{P586} has described how 20 participants were asked to enter metadata for a weather station sensor using a simple user interface. Those 20 people made 45 mistakes in total. The requirement of re-entering metadata in different places (e.g. entering metadata on GSN once and again entering metadata on OpenIoT, etc.) arises when we do not have common descriptions. Recently, the W3C Incubator Group released Semantic Sensor Network XG Final Report, which defines an SSN ontology \cite{P581}. The SSN ontology allows describing sensors, including their characteristics. This effort increases the interoperability and accuracy due to the lack of manual data entering. Furthermore, such mistakes can be avoided by letting the sensor hardware manufactures produce and make available sensor descriptions using ontologies so that IoT solution developers can retrieve and incorporate (e.g. mapping) them in their own software system.

Based on the arguments above, ontology based sensor description and data modelling is useful for IoT solutions. This approach also allows semantic querying. Our proposed solution allows the users to express their priorities in terms of sensor characteristics and it will search and select appropriate sensors. In our model, both quantitative reasoning  and semantic querying techniques are employed to increase the performance of the system by utilizing the strengths of both techniques.

\textcolor{blue}{In this paper, we propose a model that can be adopted by any IoT middleware solution. Moreover, our design can be run faster using \textit{MapReduce} based techniques,  something which increases the scalability of the solution. Our contributions can be summarized as follows. We have developed an ontology based context framework for sensors in IoT which allows capturing and modelling context properties related to sensors. This information allows users to search the sensors based on context. We have designed, implemented and evaluated our proposed CASSARAM model and its performance in a comprehensive manner. Specifically, we propose a Comparative-Priority Based Weighted Index (CPWI) technique to index and rank sensors based on the user preferences. Furthermore, we propose a Comparative-Priority Based Heuristic Filtering (CPHF) technique to make the sensor search process more efficient. We also propose  a Relational-Expression based Filtering (REF) technique to support more comprehensive searching. Finally, we propose and compare several distributed sensor search mechanisms.}

\textcolor{blue}{The rest of this paper is structured as follows: In Section \ref{sec:Background}, we briefly review the literature and provide some descriptions of leading IoT middleware solutions and their sensor searching capabilities. Next, we present the problem definitions and motivations in Section \ref{sec:Problem_Definition_and_Motivation}. Our proposed solution, CASSARAM, is presented with details in Section \ref{sec:CASSARAM}. Data models, the context framework, algorithms, and architectures are discussed in this section. The techniques we developed to improve CASSARAM are presented in Section \ref{sec:CASSARAM:Improving_Efficiency}. In Section \ref{sec:Implementation}, we provide implementation details, including tools, software platforms,  hardware platforms, and the data sets used in this research. Evaluation and discussions related to the research findings are presented in Section \ref{sec:Discussion}. Finally, we present a conclusion and prospects for future research in Section \ref{sec:Conclusions and Future Work}.}

\vspace{-5pt}

\section{Background and Related Work}
\label{sec:Background}

 Ideally, IoT middleware solutions should allow the users to express what they want and provide the relevant sensor data back to them quickly without asking the users to manually select the sensors which are relevant to their requirements. Even though IoT has received significant attention from both academia and industry, sensor search and selection has not been addressed comprehensively. Specifically, sensor search and selection techniques using context information \cite{ZMP007} have  not been explored substantially. A survey on context aware computing for the Internet of Things \cite{ZMP007} has recognised sensor search and selection as a critical task in automated sensor configuration and context discovery processes. Another review on semantics for the Internet of Things \cite{P623} has also recognised resource (e.g., a sensor or an actuator) search and discovery functionality as one of the  most important functionalities that are required in IoT. Barnaghi et al. \cite{P623} have highlighted the need for semantic annotation of IoT resources and services. Processing and analysing the semantically  annotated data are essential elements to support search and discovery \cite{P623}. This justifies our approach of annotating the sensors with related context information and using that to search the sensors. The following examples show how  existing IoT middleware solutions provide sensor searching functionality.

Linked Sensor Middleware (\textit{LSM}) \cite{P583,P584} provides some sensor selection and searching functionality. However, they have very limited capabilities, such as selecting sensors based on location and sensor type. All the searching needs to be done using SPARQL, which is not user-friendly to non-technical users. Similar to \textit{LSM}, there are several other IoT middleware related projects under development at the moment. \textit{GSN} \cite{P022}  is a platform aiming at providing flexible middleware to address the challenges of sensor data integration and distributed query processing. It is a generic data stream processing engine. \textit{GSN} has gone beyond the traditional sensor network research efforts such as routing, data aggregation, and energy optimisation. \textit{GSN} lists all the available sensors in a combo-box which users need to select. However, GSN lacks semantics to model the metadata. Another approach is Microsoft \textit{SensorMap} \cite{P578}. It only allows users to select sensors by using a location map, by sensor type and by keywords. \textit{xively} (xively.com) is also another approach which provides a secure, scalable platform that connects devices and products with applications to provide real-time control and data storage. This also provides only keyword search. The illustrations of the search functionalities provided by the above mentioned IoT solutions are presented in \cite{ZMP006}. Our proposed solution CASSARAM can be used to enrich all the above mentioned IoT middleware solutions with a comprehensive sensor search and selection functionality.

 In the following, we briefly describe some of the work done in sensor searching and selection. Truong et al. \cite{P587} propose a fuzzy based similarity score comparison sensor search technique to compare the output of a given sensor with the outputs of several other sensors to find a matching sensor. Mayer et al. \cite{P588} considers the location of smart things/sensors as the main context property and structures them in a logical structure. Then, the sensors are searched by location using tree search techniques. Search queries are distributively processed in different paths/nodes of the tree. Elahi et al. \cite{P589} propose a content-based sensor search approach (i.e. finding a sensor that outputs a given value at the time of a query). \textit{Dyser} is a search engine proposed by Ostermaier et al. \cite{P590} for real-time Internet of Things, which uses statistical models to make predictions about the state of its registered objects (sensors). When a user submits a query, \textit{Dyser} pulls the latest data to identify the actual current state to decide whether it matches the user query. Prediction models help to find matching sensors with a minimum number of sensor data retrievals. Very few related efforts have focused on sensor search based on context information.
Perera et al. \cite{ZMP006} have compared the similarities and differences between sensor search and web service search. It was found that context information has played a significant role in web service search (especially towards  web services composition). According to a study in Europe \cite{P625}, there are over 12,000 working and useful Web services on the Web. Even in such conditions, choice between alternatives (depending on context properties) has become a challenging problem. The similarities strengthen the argument that sensor selection is an important challenge at the same level of complexity as web services. On the other hand, the differences show that sensor selection will become a much more complex challenge  over the coming decade due to the scale of the IoT.

De et al. \cite{P622} have proposed a conceptual architecture, an IoT platform, to support real-world and digital objects. They have presented several semantic ontology based models that allow capturing information related to IoT resources (e.g. sensors, services, actuators). However, they are not focused on sensors and the only context information considered is location. In contrast, CASSARAM narrowly focuses on sensors and considers a comprehensive set of context information (see Section \ref{sec:OA:Context-aware_QoS_Matrix}). Guinard et al. \cite{P616} have proposed a web service discovery, query, selection, and ranking approach using context information related to the IoT domain. Similarly, TRENDY \cite{P618} is a registry-based service discovery protocol based on CoAP (Constrained Application Protocol) \cite{P624} based web services with context awareness. This protocol has been proposed to be used in the Web of Things (WoT) domain with the objective of dealing with a massive number of  web services (e.g. sensors wrapped in web services). Context information such as  hit count, battery, and response time are used to select the services. An interesting proposal is by Calbimonte et al. \cite{P619}, who have proposed an ontology-based approach for providing data access and query capabilities to streaming data sources. This work allows the users to express their needs at a conceptual level, independent of implementation. Our approach, CASSARAM, can be used to complement their work where we support context based  sensor search and they provide access to semantically enriched sensor data. Furthermore, our evaluation results can be used to understand the scalability and computational performance of their working big data paradigm as both approaches use the SSN ontology. Garcia-Castro et al. \cite{P620} have defined a core ontological model for Semantic
sensor web infrastructures. It can be used to model sensor networks (by extending the SSN ontology), sensor data sources, and the web services that expose the data sources. Our approach can also be integrated into the \textit{uBox} \cite{P617} approach, to search \textit{things} in the WoT domain using context information. Currently, \textit{uBox} performs searches based on location tags and object (sensor) classes (types) (e.g. hierarchy local/class/actuator/light).

The following table summarises the different research efforts that have addressed the challenge of sensor search. Table \ref{Tbl:Summarized Sensor_Search} lists the efforts and the number of sensors used in their experiments.

 %%%%%%%%%%%%%%%%%%%%%%%%%%%%%%%%%%%
\begin{table}[h]
\centering
\footnotesize
\renewcommand{\arraystretch}{1.15}

\caption{Number of sensors used in experimental evaluations of different sensor search approaches}
\vspace{-0.2cm}
\begin{tabular}{ m{3.5cm}  r }
\hline
Approach & Number of sensors used in experiments  \\ \hline \hline

Truong et al. \cite{P588} & 42 \\ 
Elahi et al. \cite{P589}  & 250 \\ 
Ostermaier et al. \cite{P590} & 385 \\ 
Mayer et al. \cite{P588} & 600 \\ 
Calbimonte et al. \cite{P467}\footnote{Not necessarily sensor search, but sensor data search based on sensors} & 1400 \\ 
\textit{LSM} \cite{P584} & 100,000 \\ 
\hline
\end{tabular}
\label{Tbl:Summarized Sensor_Search}
%\vspace{-0.6cm}
\end{table}
%%%%%%%%%%%%%%%%%%%%%%%%%%%%%%%%%%%

\vspace{-8pt}

\section{Problem Definition and Motivation}
\label{sec:Problem_Definition_and_Motivation}

The problem that we address in this paper can be defined as follows. Due to the increasing number of sensors available, we need to search and select sensors that provide data which will help to solve the problem at hand in the most efficient and effective way. Our objective is not to solve the users problems, but to help them to collect sensor data. The users can further process such data in their own ways to solve their problems. In order to achieve this, we need to search and select sensors  based on different pieces of context information. Mainly, we identify two categories of requirements: point-based requirements (non-negotiable) and proximity-based (negotiable) requirements. We examined the problem in detail  in \cite{ZMP006} by providing real world application scenarios and challenges. 

First, there are the point-based requirements that need be definitely fulfilled. For example, if a user is interested in measuring the temperature in a certain location (e.g. Canberra), the result (e.g. the list of sensors) should only contain sensors that can measure temperature. The user cannot be satisfied by being providing with any other type of sensor (e.g. pressure sensors). There is no  bargain or compromise in this type of requirement. Location can be identified as a point-based requirement. The second is proximity-based requirements that need be fulfilled in the best possible way. However, meeting the exact user requirement is not required. Users may be willing to be satisfied with a slight difference (variation). For example,  the user has the same interest as before. However, in this situation, the user imposes proximity-based requirements in addition to their point-based requirements. The user may request sensors having an accuracy of around 92\%, and reliability 85\%. Therefore, the user gives the highest priority  to these characteristics. The user will accept sensors that closely fulfil these requirements even though all other characteristics may not be favourable (e.g. the cost of acquisition may be high and the sensor response may be slow). It is important to note that users may not be able to provide any specific value, so the system should be able to understand the user's priorities and provide the results accordingly, by using comparison techniques.

Another motivation behind our research are statistics and predictions that show rapid growth in sensor deployment related to the IoT and Smart Cities. It is estimated that today there about 1.5 billion Internet-enabled PCs and over 1 billion Internet-enabled  mobile phones. By 2020, there will  be 50 to 100 billion devices connected to the Internet \cite{P029}. Furthermore, our work is motivated by the increasing trend of IoT middleware solutions development. Today, most of the leading midddleware solutions provide only limited sensor search and selection functionality, as mentioned in Section \ref{sec:Background}.

\textcolor{blue}{We highlight the importance of sensor search functionality using current and potential applications. Smart agriculture \cite{ZMP004} projects such as Phenonet \cite{P412} collects data from thousands of sensors. Due to heterogeneity, each sensor may have different context values, as mentioned in Section \ref{sec:OA:Context-aware_QoS_Matrix}. Context information can be used to selectively select sensors depending on the requirements and situations. For example, CASSARAM helps to retrieve data only from sensors which have more energy remaining when alternative sensors are available. Such action helps to run the entire sensor network for a much longer time without reconfiguring and recharging. The sensing as a service \cite{ZMP008} architectural model envisions an era where sensor data will be published and sold through the cloud. Consumers (i.e., users) will be allowed to select a number of sensors and retrieve data for some period as specified in an agreement by paying a fee. In such circumstances, allowing consumers to select the sensors they want based on context information is critical. For example, some consumers may be willing to pay more for highly accurate data (i.e., highly accurate sensors) while others may be willing to pay less for less accurate data, depending on their requirements, situations, and preferences.}

%\vspace{-5pt}

\section{Context-aware Sensor Search, Selection and Ranking Model}
\label{sec:CASSARAM}
In this section, we present the proposed sensor selection approach step by step in detail. First, we provide a high-level overview of the model, which describes the overall execution flow and critical steps. Then, we explain how user preferences are captured. Next, the data representation model and proposed extensions are presented. Finally, the techniques of semantic querying and quantitative reasoning are discussed with the help of some algorithms. All the algorithms presented in this paper are self-explanatory and  the common algorithmic notations used in this paper are presented in Table \ref{Tbl:Summarized taxonmy}.

%%%%%%%%%%%%%%%%%%%%%%%%%%%%%%%%%%%
\begin{table}[t!]
\centering
\footnotesize
\renewcommand{\arraystretch}{1.15}

\caption{Common Algorithmic Notation Table}
\vspace{-0.2cm}
\begin{tabular}{ c  m{6.5cm} }
\hline
Symbol & Definition  \\ \hline \hline

$\mathbb{O}$ & \textit{Ontology} consists of sensor descriptions and context property values related to all sensors  \\ 
$\mathbb{P}$ & \textit{UserPrioritySet} contains user priority value for all  context properties  \\ 
$\mathbb{Q}$ & \textit{Query} consists of point-based requirements expressed in SPARQL  \\ 
${N} / {N_{All}}$ &   Number of sensors required by the user / Total number of sensors available \\
$\mathbb{S}_{Filtered}$ & This contains the results of the query $\mathbb{Q}$   \\
$\mathbb{S}_{Results}$ & \textit{ResultsSet} contains selected number of sensors  \\
$\mathbb{S}_{Indexed}$ & \textit{IndexedSensorSet} store the index values of the sensors \\
$\mathbb{M}$ & Multidimensional space where each context property is represented by a dimension and sensors are plotted \\
$\mathbb{UI}$ & \textit{UserInput} consists of input values provided by the users via the user interface \\
$\mathbb{SC}/{SC}$ &  Values of all the sliders / Value of a slider  \\ 
$\mathbb{P}^{w}$ & This contains user priority value converted into weights using normalization\\
${p}_{i} / {p}_{i}^{w}$ & Value of $i^{th}$ context property / Value of $i^{th}$ context property in normalized form\\
$\mathbb{CP} / {CP}$ & \textit{ContextPropertySet} consists of all context information / value of $i$$^{th}$ context property\\
$\mathbb{NCP}$ & NormalizedContextPropertySet \\
${M}$ & Margin of error \\
$S_{j}$ & This is the $j^{th}$ sensor \\
$CP_{i}^{S_{j}}$ & {CP} value of $i^{th}$ property  of $j^{th}$ sensor. \\
${CP}^{ideal}$ & {CP} values of the ideal sensors that user prefers\\
\hline
\end{tabular}
\label{Tbl:Summarized taxonmy}
%\vspace{-0.6cm}
\end{table}
%%%%%%%%%%%%%%%%%%%%%%%%%%%%%%%%%%%

\begin{figure}[b]
 \centering
 \vspace{-0.23cm}
 \includegraphics[scale=.84]{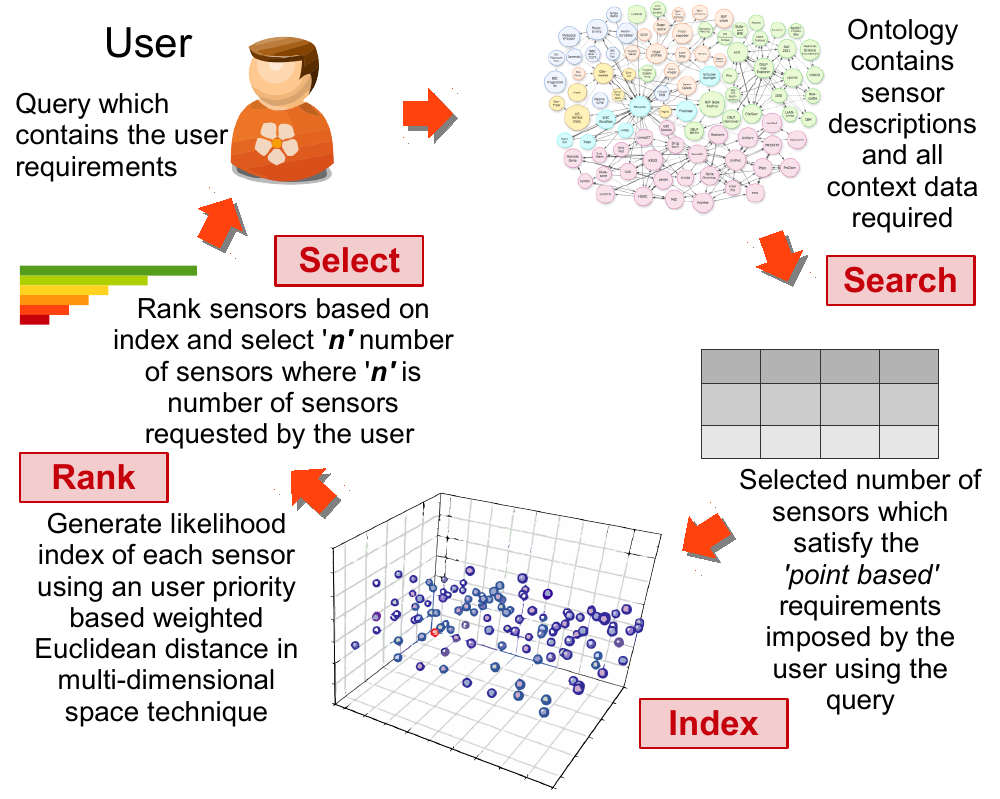}
\vspace{-0.33cm}	
 \caption{\footnotesize High level Overview of CASSARAM}
 \label{Figure:System_Architecture}	
%\vspace{-0.63cm}	
\end{figure}

\subsection{High-level Model Overview}
\label{sec:High-level_System_Architecture}

The critical steps of CASSARAM are presented in Fig. \ref{Figure:System_Architecture}. As we mentioned earlier our objective is to allow the users to search and select the sensors that best suit their requirements. In our model, we divide user requirements into two categories (from the user's perspective): \textit{point-based requirements} and \textit{proximity-based requirements}, as discussed in Section \ref{sec:Problem_Definition_and_Motivation}. Algorithm \ref{Alg:System_Flow_of_CASSARA_Model} describes the execution flow of CASSARAM. At the beginning, CASSARAM identifies the point-based requirements, the proximity-based requirements, and the user priorities. First, users need to select the point-based requirements. For example, a user may want to collect sensor data from 1,000 temperature sensors deployed in Canberra. In this situation, the sensor type (i.e., temperature), location (i.e., Canberra) and number of sensors required (i.e., 1,000) are the point-based requirements. Our CASSARAM prototype tool  provides a user interface to express this information via SPARQL queries. In CASSARAM, any context property can become a point-based requirement. Next,  users can define the proximity-based requirements. All the context properties we will present in Section \ref{sec:OA:Context-aware_QoS_Matrix} are available to be defined in comparative fashion by setting the priorities via a slider-based user interface, as depicted in Fig. \ref{Figure:User_Priority_Capturing}. Next, each sensor is plotted in a multi-dimensional space where each dimension represents a context property (e.g. accuracy, reliability, latency). Each dimension is normalized [0,1] \textcolor{black}{ as explained in Algorithm \ref{Alg:Flexi-Dynamic_Normalization}.} Then, the Comparative-Priority Based Weighted Index (CPWI) is generated for each sensor by combining the user's priorities and context property values as explained in Section \ref{sec:OA:Statistical_Reasoning}. The sensors are ranked using the CPWI and the number of sensors required by the user is selected from the top of the list.

%\vspace{-6pt}

 \begin{algorithm}[t]

 \algsetup{linenosize=}
 \caption{Execution Flow of CASSARAM}
 \label{Alg:System_Flow_of_CASSARA_Model}
 \begin{algorithmic}[1]

 \REQUIRE ($\mathbb{O}$), ($\mathbb{P}$), ($\mathbb{Q}$),   (${N}$),  ($\mathbb{M}$).  
 \STATE \textbf{Output:}  $\mathbb{S}_{Results}$
 \STATE $\mathbb{S}_{Filtered} \leftarrow \textrm{queryOntology}(\mathbb{O},\mathbb{Q})$
 \IF{$\textrm{cardinality}(\mathbb{S}_{Filtered})< {N}$ }
 \RETURN $\mathbb{S}_{Results}  \leftarrow    \mathbb{S}_{Filtered} $
 \ELSE
 \STATE $\mathbb{P}  \leftarrow \textrm{capture user priorities}(\mathbb{UI})$ 
 \STATE $\mathbb{M}  \leftarrow \textrm{Plot sensors in multidimensional space} (\mathbb{S}_{Results})$
 \STATE $\mathbb{S}_{Indexed} \leftarrow \textrm{calculate CPWI} ({\mathbb{S}_{Results},\mathbb{M}})$
 \STATE $\mathbb{S}_{Results}  \leftarrow \textrm{rank sensors}(\mathbb{S}_{Indexed})$
  \STATE $\mathbb{S}_{Results} \leftarrow \textrm{select sensors}(\mathbb{S}_{Results},N)$
 \RETURN $\mathbb{S}_{Results}$
 \ENDIF

 \end{algorithmic}
%\vspace{-10pt}
 \end{algorithm}

\vspace{-10pt}

\subsection{Capturing User Priorities}
\label{sec:User_Priority_Capturing}

\begin{figure}[t]
 \centering
%\vspace{-0.58cm}
 \includegraphics[scale=.80]{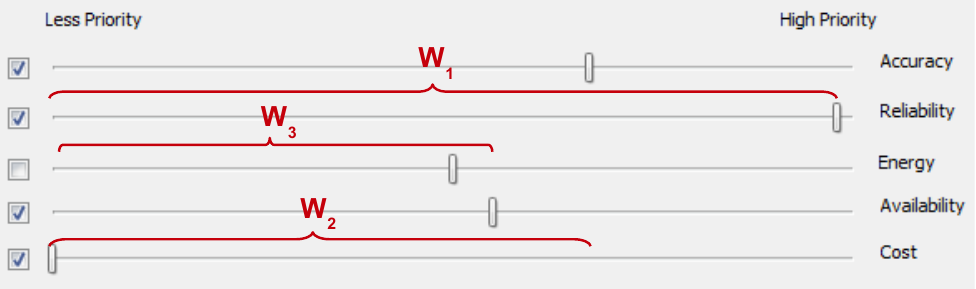}
\vspace{-0.23cm}	
 \caption{ \footnotesize A weight of $W_{1}$ is assigned to the \textit{reliability} property. A weight of $W_{2}$ is assigned to the \textit{Accuracy} property. A weight of $W_{3}$ is assigned to the \textit{availability} property. A weight of $W_{4}$, the default weight, is assigned to the \textit{cost} property. High priority means always favoured, and low priority means always disfavoured. For example, if the user makes \textit{cost} a high priority (more towards the right), that means CASSARAM tries to find the sensors that produce data at the lowest cost. Similarly, if the user makes \textit{accuracy} a high priority, that means CASSARAM tries to find the sensors that produce data with high accuracy.}
 \label{Figure:User_Priority_Capturing}	
\vspace{-0.58cm}	
\end{figure}

This is a technique we developed to capture the user's priorities through a user interface, as shown in Fig. \ref{Figure:User_Priority_Capturing}. CASSARAM allows users to express which context property is more important to them, when compared to others. If a user does not want a specific context property to be considered in the indexing process, they can avoid it by not selecting the check-box correlated with that specific context property. For example, according to Fig. \ref{Figure:User_Priority_Capturing}, \textit{energy} will not be considered when calculating the CPWI. This means the user is willing to accept sensors with any \textit{energy} consumption level. Users need to position the slider of each context property if that context property is important to them. The slider scale begin from 1, which means  no priority (i.e., the left corner). The highest priority can be set by the user as necessary with the help of a \textit{scaler}, where a higher scale makes the sliders more sensitive (e.g. $10^{2}= 1 \textrm{ to } 100, 10^{3}, 10^{4}$). Algorithm \ref{Alg:User_Priority_Capturing} describes the user priority capturing process.

 \begin{algorithm}[b]
 \algsetup{linenosize=}
 \caption{User Priority Capturing}
 \label{Alg:User_Priority_Capturing}
 \begin{algorithmic}[1]
 \REQUIRE ($\mathbb{UI}$), ($\mathbb{SC}$)
 \STATE \textbf{Output:} $\mathbb{P}^{w}$
 \STATE $\mathbb{P} \leftarrow \textrm{extract user priorities}(\mathbb{UI})$
 \STATE $SC_{Highest} \leftarrow \textrm{get maximum priority}(\mathbb{SC})$
 \STATE $SC_{Lowest} \leftarrow \textrm{get minimum priority}(\mathbb{SC})$
  \STATE $SC_{Range} \leftarrow SC_{Highest} - SC_{Lowest}$
 \FOR {each context property priority $p_{i}$ $\in \mathbb{P} $}
 \STATE $p_{i}^{w} \leftarrow  (p_{i} \div S_{Range})$
  \IF{$ p_{i}^{w} \geq 0$ }
  \STATE add $p_{i}^{w}$ to $\mathbb{P}^{w}$
  \ELSE 
  \STATE continue
  \ENDIF 
 \ENDFOR
 \RETURN $\mathbb{P}^{w}$ 
 \end{algorithmic}
 \end{algorithm}

As depicted in Fig. \ref{Figure:User_Priority_Capturing}, if the user wants more wieght to be placed on the reliability of a sensor than on its accuracy, the reliability slider need to be placed further to the right than the accuracy slider. A weight is calculated for each context property. Therefore, higher priority means higher weight. As a result, sensors with high reliability and accuracy will be ranked highly. However, those sensors may have high costs due to the low priority placed on cost.

\vspace{-5pt}

\subsection{Data Modelling and Representation}
\label{sec:Data_Modelling_and_Representation}

\begin{figure*}[t]
 \centering

 \includegraphics[scale=.57]{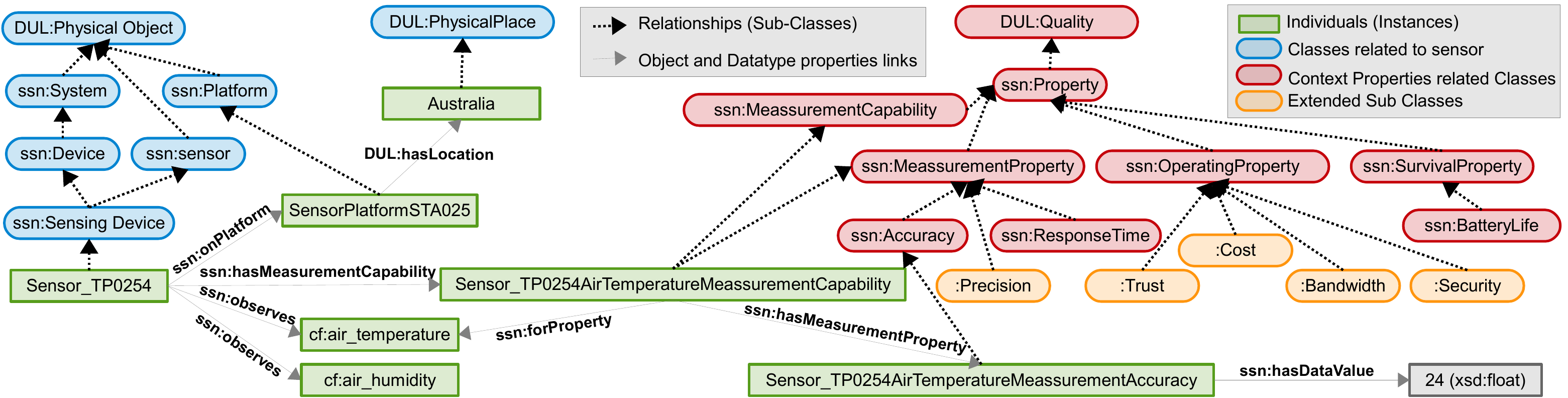}
\vspace{-0.53cm}	
 \caption{\footnotesize Data model used in CASSARAM. In SSN ontology, \textit{sensors} are not constrained to physical sensing devices; rather a sensor is anything that can estimate or calculate the value of a phenomenon, so a device or computational process or combination could play the role of a sensor. A \textit{sensing device} is a device that implements sensing  \cite{P581}. \textit{Sensing device} is also a sub class of \textit{sensor}. By following above definition, our focus is on sensors. CF (Climate and Forecast) ontology is a domain specific external ontology. DOLCE+DnS Ultralite (DUL) ontology provides a set of upper level concepts that can be the basis for easier interoperability among many middle and lower level ontologies. More details are provided in \cite{P581}.}
 \label{Figure:Data_Model}	
\vspace{-0.43cm}	
\end{figure*}

In this paper, we employed the Semantic Sensor Network Ontology (SSN) \cite{P581} to model the sensor descriptions and context properties. The main reasons for selecting the SSN ontology are its interoperability and the trend towards ontology usage in the IoT and sensor data management domain. A comparison of different semantic sensor ontologies is presented in \cite{P103}. The SSN ontology is capable of modelling a significant amount of information about sensors, such as sensor capabilities, performance, the conditions in which it can be used, etc. The details are presented in \cite{P581}. The SSN ontology includes the most common context properties, such as accuracy, precision, drift, sensitivity, selectivity, measurement range, detection limit, response time, frequency and latency. However, the SSN ontology can be extended \emph{unlimitedly} by a categorization with three classes: \textit{measurement property}, \textit{operating property}, and \textit{survival property}. We depict a simplified segment of the SSN ontology in Fig. \ref{Figure:Data_Model}. We extend the \textit{quality class} by adding several sub-classes based on our context framework, as listed in Section \ref{sec:OA:Context-aware_QoS_Matrix}. All  context property values are stored in the SSN ontology in their original measurement units. CASSARAM normalizes them on demand to [0,1] to ensure consistency. Caching techniques can be used to increase the  execution performances. Due to technological advances in sensor hardware development, it is impossible to statically define upper and lower bounds for some context properties (e.g. battery life will be improved over time due to advances in sensor hardware technologies). Therefore, we propose Algorithm \ref{Alg:Flexi-Dynamic_Normalization} to dynamically normalize the context properties.

\vspace{-10pt}

\subsection{Filtering Using Querying Reasoning}
\label{sec:OA:Semantic_Reasoning}

Once the point-based requirements of the user have been identified, they need to be expressed using SPARQL. Semantic querying has weaknesses and limitations. When a query becomes complex, the performance decreases \cite{P591}. Relational expression based filtering can also be used; however,  using it will increase the computational requirements. Further explanations are presented in Section \ref{sec:OA:Improving Efficiency}. Any of the context properties identified in Section \ref{sec:OA:Context-aware_QoS_Matrix} can become point-based requirements and need to be represented in SPARQL. This step produces  $\mathbb{S}_{Filtered} $, where all the sensors satisfy all the point-based requirements.

 \begin{algorithm}[t]
 \algsetup{linenosize=}
 \caption{Flexi-Dynamic Normalization}
 \label{Alg:Flexi-Dynamic_Normalization}
 \begin{algorithmic}[1]
 \REQUIRE ($\mathbb{CP}$), ($\mathbb{S}$), ($cp_{i}$), 
  
 \STATE \textbf{Output:} $\mathbb{NCP}$
 \STATE $cp_{i}^{S_{j}} \leftarrow $ receive new property value$^{*}$ 
 \STATE $cp_{i}^{highest} \leftarrow \textrm{retrieve highest}(\mathbb{CP})$
  \STATE $cp_{i}^{lowest} \leftarrow \textrm{retrieve lowest}(\mathbb{CP})$
 
 \IF { $cp_{i}^{highest} <  cp_{i}^{S_{j}} $ }
 \STATE $cp_{i}^{highest}\leftarrow cp_{i}^{S_{j}}$
  \FOR {each $cp_{i}^{S_{j}} \in \mathbb{CP,S} $ }
   \STATE $\textrm{update}(\mathbb{NCP})   \leftarrow [\frac{(cp_{i}^{S_{j}} - cp_{i}^{lowest})}{(cp_{i}^{highest} - cp_{i}^{lowest})}]$ 
 \ENDFOR
 \ELSE
 \STATE $\textrm{update}(\mathbb{NCP})   \leftarrow [\frac{(cp_{i}^{S_{j}} - cp_{i}^{lowest})}{(cp_{i}^{highest} - cp_{i}^{lowest})}]$ 
 \ENDIF
 \RETURN $\mathbb{NCP}$
 
 $^{*}$\footnotesize sensors registered in the IoT middleware
 \end{algorithmic}
 \end{algorithm}

%\vspace{-25pt}

\subsection{Ranking Using Quantitative Reasoning}
\label{sec:OA:Statistical_Reasoning}

In this step, the sensors are ranked based on the proximity-based user requirements. We developed a weighted Euclidean distance based indexing technique, called the Comparative-Priority Based Weighted Index (CPWI),  as follows.

\begin{center}
$\left ( CPWI \right ) = \sqrt{\sum_{i=1}^{n} \left [  W_{i}(U_{i}^{d} -  S_{i}^{\alpha })^{2}\right ]    } $
\end{center}

\begin{figure}[t]
 \centering
 \vspace{-0.30cm}
 \includegraphics[scale=.55]{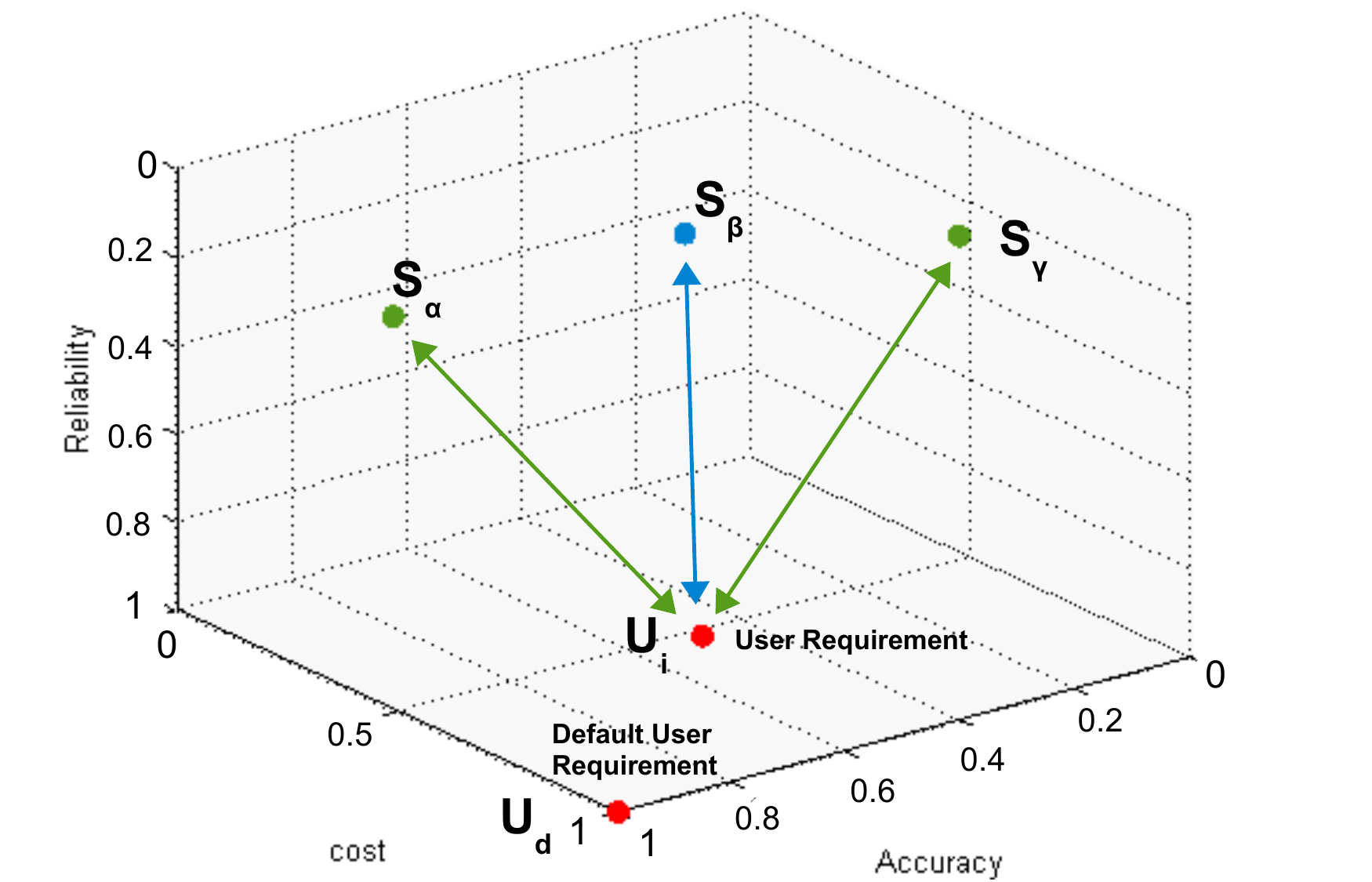}
 \vspace{-0.63cm}	
 \caption{\footnotesize Sensors plotted in three-dimensional space for demonstration purposes. $S_{\alpha}$, $S_{\beta}$, and $S_{\gamma}$ represent real sensors. $U_{i}$ represent the user preferred sensor.  $U_{d}$ represent the default user preferred sensor. CPWI calculate weighted distance between $S_{j=\alpha || \beta || \gamma }$ and $U_{i||d}$. Shortest distance means sensor will rank higher because it is close to the user requirement. }
 \label{Figure:Sensor_Plot}	
 \vspace{-0.33cm}	
 \end{figure}

First, each sensor is plotted in multi-dimensional space where each context property is represented by a dimension. Then, users can plot an ideal sensor in the multi-dimensional space by manually entering context property values as illustrated in Fig. \ref{Figure:Sensor_Plot} by $U_{i}$. By default, CASSARAM will automatically plot an ideal sensor as depicted in $U_{d}$ (i.e., the highest value for all context properties). Next, the priorities defined by the user are retrieved. Based on the positions of the sliders (in Fig. \ref{Figure:User_Priority_Capturing}), weights are calculated in a comparative fashion. Algorithm \ref{Alg:Comparative_Priority-based_Weighted_Indexing} describes the indexing process.  It calculates the CPWI and ranks the sensors using reverse-normalised techniques in descending order. CASSARAM selects ${N}$ sensors from the top.  

 \begin{algorithm}[t]
 \algsetup{linenosize=}
 \caption{Comparative-Priority Based Weighted Index}
 \label{Alg:Comparative_Priority-based_Weighted_Indexing}
 \begin{algorithmic}[1]
 \REQUIRE   ($\mathbb{P}^{w}$), ($\mathbb{CP}$), ($\mathbb{S}^{Indexed}$), ($\mathbb{P}^{S_{j}}$), ($\mathbb{UI}$)
 
 \STATE \textbf{Output:} $\mathbb{S}_{Ranked}$
  \STATE ${CP}^{ideal}\leftarrow \textrm{proximity based requirements}(\mathbb{UI})$
   \STATE $\textrm{plot on multi-dimensional space} ({CP}^{ideal})$
   \FOR { each sensor $S_{j} \in \mathbb{S}$ }
    \STATE $\textrm{plot on multi-dimensional space} ({CP}^{S_{j}})$ 
  \ENDFOR
        \STATE Indexing Formula (for $S^{\alpha}$) = $\sqrt{\sum_{i=1}^{n} \left [  W_{i}(U_{i}^{d} -  S_{i}^{\alpha })^{2}\right ]    } $
     \FOR { each sensor $s^{j} \in \mathbb{S}$ }

      \STATE $ \mathbb{S}^{Indexed} \leftarrow \textrm{calculate index} (\mathbb{P}^{S_{j}}, \mathbb{P}^{w})$ 
    \ENDFOR
 \STATE    $\mathbb{S}_{Ranked} \leftarrow \textrm{reversed normalized ranking$^{*}$}(\mathbb{S}^{Indexed}) $   {\footnotesize  $^{*}$i.e.: lowest value is ranked higher which represents the weighted distance between use preferred sensor and the real sensors}

 \RETURN $\mathbb{S}_{Ranked}$

 \end{algorithmic}
 
 \end{algorithm}

\vspace{-12pt}

\subsection{Context Framework}
\label{sec:OA:Context-aware_QoS_Matrix}

After evaluating a number of research efforts conducted in the quality of service domain relating to web services \cite{P593}, mobile computing \cite{P592}, mobile data collection \cite{ZMC008}, and sensor ontologies \cite{P581}, we extracted the following context properties to be stored and maintained in connection with each sensor. This information helps to decide which sensor is to be used in a given situation. We adopt the following definition of \textit{context} for this paper. ``Context is any information that can be used to characterise the situation of an entity. An entity is a person, place, or object that is considered relevant to the interaction between a user and an application, including the user and applications themselves.''\cite{P104}. CASSARAM has no limitations on the number of context properties that can be used. More context information can be added to the following list as necessary. Our context framework comprises availability, accuracy, reliability, response time, frequency, sensitivity, measurement range, selectivity, precision, latency, drift, resolution, detection limit, operating power range, system (sensor) lifetime, battery life, security, accessibility, robustness, exception handling, interoperability, configurability, user satisfaction rating, capacity, throughput, cost of data transmission, cost of data generation, data ownership cost, bandwidth, and trust.

\vspace{-5pt}

\section{Improving Scalability and Efficiency}
\label{sec:CASSARAM:Improving_Efficiency}

\textcolor{blue}{In this  section, we present three approaches that improve the efficiency and the capability of CASSARAM. First, we propose a heuristic approach that can handle a massive number of sensors by trading off with accuracy. Second, we propose a relational-expression based filtering  technique that saves computational resources. Third, we tackle the challenge of distributed sensor search and selection.}

%\vspace{-8pt}

\subsection{Comparative-Priority Based Heuristic Filtering (CPHF)}
\label{sec:IE:Comparative_Priority-based_Heuristic_Filtering}

\begin{algorithm}[t]
 
 \algsetup{linenosize=}
 \caption{Comparative-Priority Based Heuristic Filtering}
 \label{Alg:Comparative_Priority-based_Heuristic_Reduction}
 \begin{algorithmic}[1]
 \REQUIRE ($\mathbb{O}$), ($\mathbb{P}$), ($\mathbb{Q}$), (${N}$), ($M$\%)
  
 \STATE \textbf{Output:} $\mathbb{S}_{Filtered}$
 
 \STATE $\mathbb{S} \leftarrow \textrm{query ontology}(\mathbb{O},\mathbb{Q})$
 \STATE $\mathbb{P}^{w} \leftarrow \textrm{get weighted priorities}(\mathbb{P})$   
  \STATE $\mathbb{P}^{Percentages} \leftarrow \textrm{convert weights to percentages}(\mathbb{P}^{w})$ 
  \STATE ${N}_{All} \leftarrow \textrm{total numberof available sensors}(\mathbb{O},\mathbb{Q})$
  
    \STATE ${N} \leftarrow \textrm{required number of sensors}(\mathbb{UI})$
    
   \STATE ${N}_{Removable} \leftarrow ({N}_{All} - {N}) $
  \STATE $\mathbb{P}^{Percentages}_{ordered} \leftarrow \textrm{descending order} (\mathbb{P}^{Percentages} )$

\FOR {each priority percentage $p \in \mathbb{P}^{Percentages}_{ordered} $}
 \STATE  $\mathbb{S}_{Filtered} \leftarrow $ Query $\mathbb{S}^{Filtered}$ and ordered by $p $ 
  \STATE Remove $    {N}_{Removable} \times (100-M) $ sensors from bottom.
 \ENDFOR

 \RETURN $\mathbb{S}_{Filtered}$
 \end{algorithmic}
 \end{algorithm}

\begin{figure}[t]
\centering
\vspace{-8pt}
\includegraphics[scale=.60]{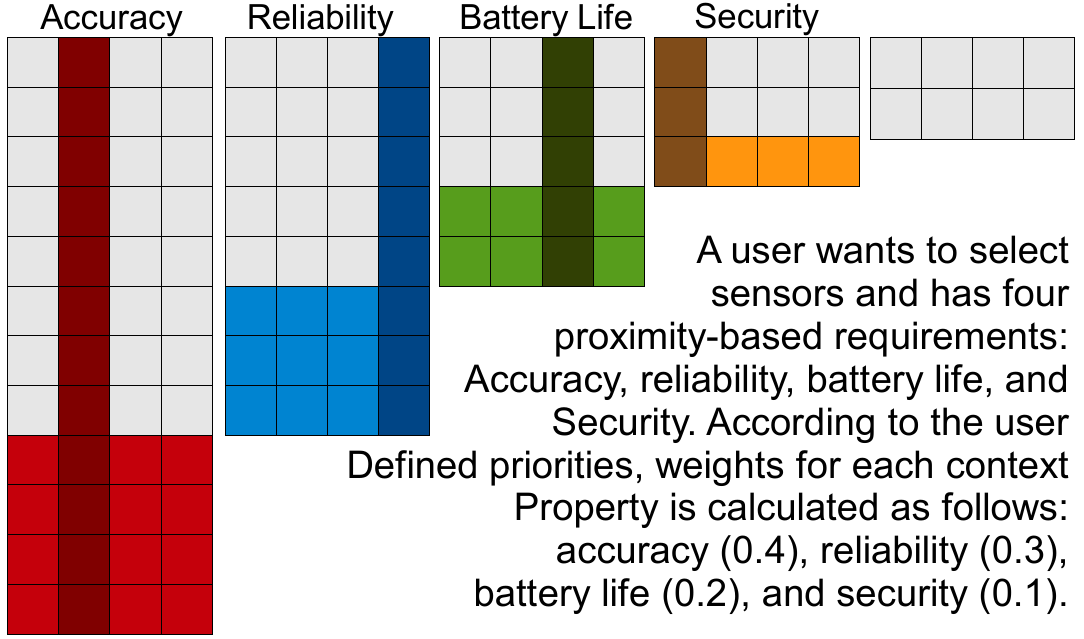}
\vspace{-0.23cm}
	
\caption{\footnotesize Visual illustration of Comparative-Priority Based Heuristic Filtering }
\label{Figure:Comparative_Priority-based_Heuristic_Filtering}	
\vspace{-0.63cm}	
\end{figure}

\textcolor{blue}{The solution we discussed so far works well with small number of sensors. However, model becomes inefficient when the number of sensors available to search increases. Let us consider an example to identify the inefficiency. Assume we have access to one million sensors. A user wants to select 1,000 sensors out of them. In such situation, CASSARAM will index and rank one million sensors using proximity-based requirements provided by the user and select top 1,000 sensors. However, indexing and ranking all possible sensors (in this case one million) is inefficient and wastes significant amount of computational resources. Furthermore, CASSARAM will not be able to process large number of user queries due to such inefficiency. We propose a technique called Comparative-Priority Based Heuristic Filtering (CPHF) to make CASSARAM more efficient. The execution process is explained in Algorithm \ref{Alg:Comparative_Priority-based_Heuristic_Reduction}.} \textcolor{blue}{The basic idea is to remove sensors that are positioned far away from user defined ideal sensor and reduce the number of sensors that need to be indexed and ranked. Fig. \ref{Figure:Comparative_Priority-based_Heuristic_Filtering} illustrates the CPHF approach with a sample scenario. The CPHF approach can be explained as follows. First, all the eligible sensors are ranked in descending order of the highest weighted context property (in this case accuracy). Then, 40\% (from ${N}_{Removable}$) of the sensors from the bottom of the list need to be removed. Next, the remaining sensors need to be ordered in descending order of the next highest weighted context property (in this case reliability). Then, 30\% (from ${N}_{Removable}$) of the sensors from the bottom of the list need to be removed. This process needs to be applied for the remaining context properties as well. Finally, the remaining sensors need to be indexed and ranked. This approach dramatically reduces the indexing and ranking related inefficiencies. Broadly, this category of techniques are called \textit{Top-K selection} where top sensors are selected in each iteration. The efficiency of this approach is evaluated and discussed in Section \ref{sec:Discussion}.}
\vspace{-1pt}

%\vspace{-10pt}

\subsection{Relational-Expression Based Filtering (REF)}
\label{sec:OA:Improving Efficiency}

\textcolor{blue}{This section explains how computational resources can be saved and how to speed up the sensor search and selection process by allowing the users to define preferred context property values using relational operators such as $<, >, \leq$, and $\geq$. For example, users can define an upper bound, lower bound, or both, using relational operators. All context properties defined by relational operators, other than the equals sign ($=$), are considered to be semi-non-negotiable requirements. According to CASSARAM, non-negotiable as well as semi-non-negotiable requirements are defined using semantic queries. Let us consider a scenario where a user wants to select sensors that have 85\% accuracy. However, the user can be satisfied by providing sensors with accuracy between 70\% and 90\%. Such requirements are called semi-non-negotiable requirements. Defining such a range helps to ignore irrelevant sensors during the semantic querying phase without even retrieving them to the CPWI generating phase, and this saves computational resources. Even though users may define ranges, the sensors will be ranked  considering the user's priorities by applying the same concepts and rules as explained in Section \ref{sec:CASSARAM}. The efficiency of this approach is evaluated in Section \ref{sec:Discussion}.}

%\vspace{-0.30cm}

\subsection{Distributed Sensor Searching}
\label{sec:CASSARAM:Context-aware_Distributed_Searching}

\begin{figure}[t]
\centering
% \vspace{-0.43cm}	
\includegraphics[scale=.70]{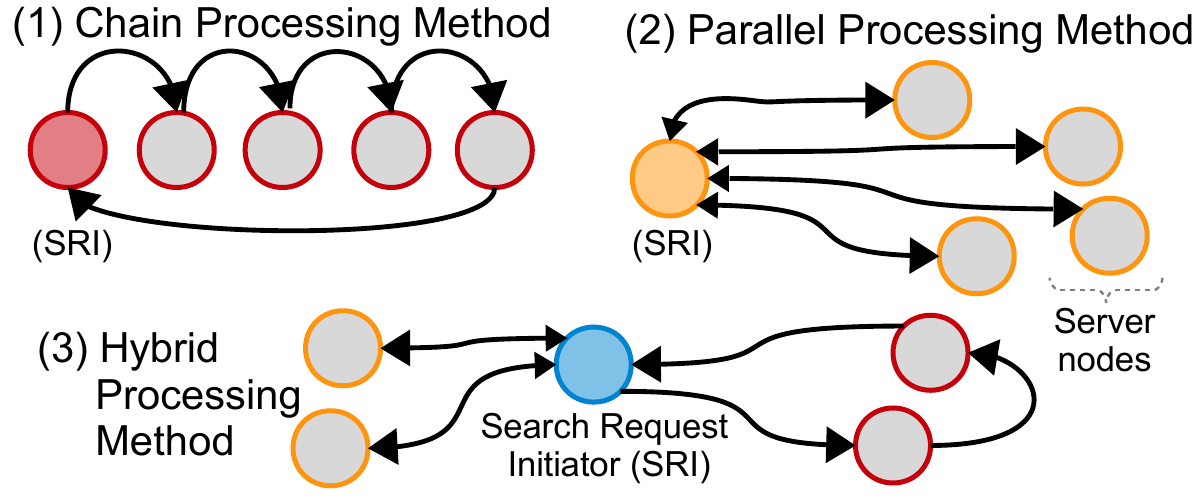}
\vspace{-0.23cm}
	
\caption{\footnotesize Distributed Processing Approaches for CASSARAM}
\label{Figure:Distrbuted_Processing}	
\vspace{-0.43cm}	
\end{figure}

\textcolor{blue}{We have explained how CASSARAM works in an isolated environment without taking into consideration the distributed nature of the problem. Ideally, we expect that not all sensors will be connected to one single server (e.g., a single middleware instance). Similarly, it is extremely inefficient to store complete sensor descriptions and related context information in many different servers in a redundant way. Ideally, each IoT middleware instance should keep track of the sensors that are specifically connected to them. This means that each server knows only about a certain number of sensors. However, in order to deal with complex user requirements, CASSARAM may need to query multiple IoT middleware instances to search and select the suitable sensors. Let us consider a scenario related to the smart agriculture domain \cite{ZMP004}. A scientist wants to find out whether his experimental crops have been infected with a disease. His experimental crops are planted in fields distributed across different geographical locations in Australia. Furthermore, the sensors deployed in the fields are connected to different IoT middleware instances, depending on the geographical location. In order to help the user to find the appropriate sensors, CASSARAM needs to query different servers in a distributed manner. We explored the possibilities of performing such distributed queries efficiently. We identified three different ways to search sensors distributively, depending on how the query/data would be transferred over the network (i.e., path), as depicted in Fig. \ref{Figure:Distrbuted_Processing}. We also identified their strengths, weaknesses, and applicability to different situations.}

\textcolor{blue}{\textit{1) Chain Processing:} Data is sent from one node to another sequentially as depicted in Fig. \ref{Figure:Distrbuted_Processing}(a). First, a user defines his requirements using an IoT middleware instance (e.g. GSN installed in a particular server). Then, this server becomes the search request initiator (SRI) for that specific user request. The SRI processes the request and selects the 100 most appropriate sensors. Then, the information related selected sensors (i.e. the unique IDs of the sensors and respective CPWIs) is sent to the next server node. The second node (i.e., that next node) merges the incoming sensor information with the existing sensor descriptions and performs the sensor selection algorithm and selects the 100 best sensors. This pattern continues until the sensor request has visited all the server nodes. This method saves communication bandwidth by transferring only the most essential and minimum amount of data. In contrast, due to a lack of parallel processing, the response time could be high.}

\textcolor{blue}{\textit{2) Parallel Processing:} The SRI parallelly sends each user search request to all available nodes. Then, each sensor node performs the sensor searching algorithm at the same time. Each node selects the 100 most appropriate sensors and returns the information related selected sensors to the SRI. In circumstances where we have 2500 server nodes, the amount of data ($2500 \times 100$) received by the SRI could be overwhelming, which would waste the communication bandwidth. The SRI processes the sensor information ($2500 \times 100$) and selects the final 100 most appropriate sensors. This approach becomes inefficient when ${N}$ becomes larger.}

\textcolor{blue}{\textit{3) Hybrid Processing:} By observing the characteristics of the previous two methods, it is obvious that the optimal distributed processing strategy should employ both chain and parallel processing techniques. There is no single method that works efficiently for all types of situations. An ideal distributed processing strategy for each situation needs to be designed and configured dynamically depending on the context, such as the types of the devices, their capabilities, bandwidth available, and so on.}

\begin{figure}[t]
\centering
%\vspace{-0.53cm}	
\includegraphics[scale=.77]{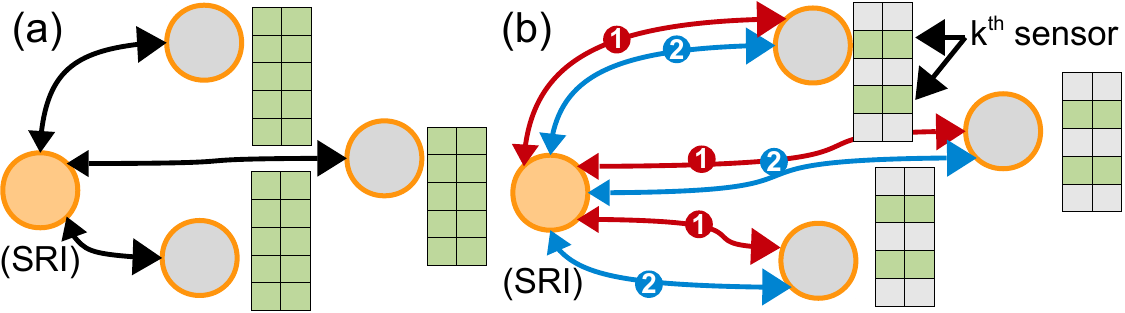}
\vspace{-0.53cm}
	
\caption{\footnotesize Optimization: (a) wihout $k$-extension and (b) with $k$-extension.}
\label{Figure:Distrbuted_Processing_Optimization}	
\vspace{-0.43cm}	
\end{figure}

\textcolor{blue}{We can improve the efficiency of the above methods as follows. In the parallel processing method, each node sends information related to ${N}$ sensors to the SRI as depicted in Fig. \ref{Figure:Distrbuted_Processing_Optimization}(a). However, at the end, the SRI may only select ${N}$ sensors (in total) despite its having received a significant amount of sensor related information  (${N} \times number of nodes$). Therefore, the rest of the data $[({N} \times number of nodes)-{N}]$ received by the SRI would be wasted.  For example, let us assume that a user wants to select 10,000 sensors. Assuming that there are 2500 server nodes, the SRI may receive a significant amount of sensor information ($10,000 \times 2500$). However, it may finally select only 10,000 sensors. We propose the following method to reduce this wastage, depicted in Fig. \ref{Figure:Distrbuted_Processing_Optimization}(b). }

\textcolor{blue}{In this method, the SRI forwards the search request to each server node parallelly, as depicted in step (1) in Fig. \ref{Figure:Distrbuted_Processing_Optimization}b. Each node selects the 10,000 most appropriate sensors. Without sending information about these 10,000 sensors to the SRI, each server node sends only information about the $k$th sensor (the UID and CPWI of every $k$th sensor). (I.e., If $k=1,000$, then the server node sends only the 1000th, 2000th, 3000th, \dots 10,000th sensors). Therefore, instead of sending 10,000 records, now each server node returns only 10 records. Once the SRI receives the sensor information from all the server nodes, it processes and decides which portions need to be retrieved. Then, the SRI sends requests back to the server nodes and now each node returns the exact portion specified by the SRI (e.g. the 5th server node may return only the first 2000 sensors instead of sending 10,000 sensors) as depicted in (2). In this method, $k$ plays a key role and has a direct impact on the efficiency. $k$ needs to be chosen by considering ${N}$ as well as other relevant context information as mentioned earlier. For example, if we use a smaller $k$, then information about more sensors would be sent to the SRI during step (1), but with less wastage in step (2). In contrast, if  we use a larger $k$, then less information would be sent to the SRI during step (1), but there would be comparatively more wastage in step (2). Furthermore, machine learning techniques can be used to customize the value of $k$ for each server node, depending on the user's request and context information, such as the types of the sensors, energy, bandwidth availability, etc. The suitability of each approach is discussed in Section \ref{sec:E:Evaluating_Distributed_Sensor_Searching}.}

\vspace{-0.13cm}

\section{Implementation and Experimentation}
\label{sec:Implementation}

%\vspace{-2pt}

\begin{figure*}[h]
\centering
\includegraphics[scale=1.0]{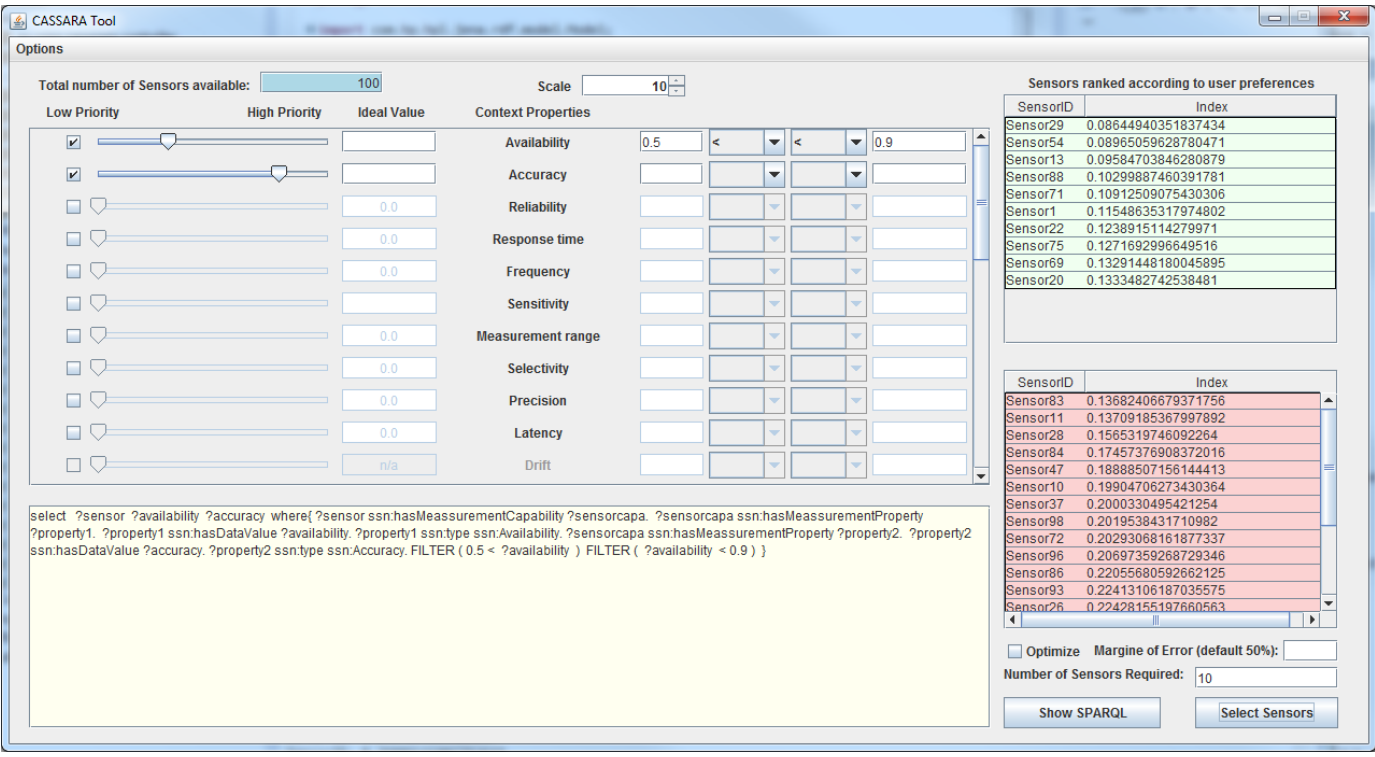}
\vspace{-0.13cm}	
\caption{ \footnotesize First, users need to select, in the UI, the context properties about which they are concerned. Then, users need to set the scale. The slider becomes more sensitive when the scale is increased. Next, the slider attached to each context property needs to be positioned to express its priority. The ideal value related to each context property can be entered. The values can be entered in native measurement units (e.g., accuracy in percentage, latency in milliseconds). All the values are normalized by CASSARAM. The default is \textit{`best possible'} (i.e., highest accuracy, lowest cost, lowest latency). Later, users can decide whether to use the optimization functionality or not, by selecting that option. Users can also define the margin of error as a percentage (the default is 50\%). Based on the user's preferences, CASSARAM generates the SPARQL appropriately. Finally, users need to specify the number of sensors they require.}
\label{Figure:CASSARA_Tool_GUI}	
\vspace{-0.43cm}	
\end{figure*}

In this section, we describe the experimental setup, datasets used, and assumptions. The experimental scenarios we used are explained at the end. The discussions related to the experiments are presented in Section \ref{sec:Discussion}.

We analysed and evaluated the proposed model using a prototype, called \textit{`CASSARA Tool'}, which we developed using Java.  The user interface of \textit{`CASSARA Tool'} is presented in Fig. \ref{Figure:CASSARA_Tool_GUI} with a self-explanatory description. The data was stored in a MySQL database. Our tool allows capturing user preferences and the priorities of the various context properties of the sensors. We used a computer with an Intel(R) Core i5-2557M 1.70GHz CPU  and 4GB RAM to evaluate our proposed model. We also reproduced  the experimentations using a higher-end computer with more CPU and RAM and the results showed that the graphs are similar in shape though the exact values are different. In order to perform mathematical operations such as a Euclidean distance calculation in multi-dimensional space, we used  the Apache Commons mathematics \cite{P580} library. It is an open source optimized library of lightweight, self-contained mathematics and statistics components, addressing the most common problems not available in the Java programming language. As we used a Semantic Sensor Ontology (SSN) \cite{P581} to manage the sensor descriptions and related data, we employed open source Apache Jena API \cite{P436} to process and manipulate the semantic data. Our evaluation used a combination of real data and synthetically generated data. We collected environmental linked data from the Bureau of Meteorology \cite{P582} and data sets from  both the Phenonet project \cite{P412}  and  the Linked Sensor Middleware (LSM) project \cite{P583,P584}. The main reasons for combining the data were the need for a  large amount of data  and the need to control different aspects (e.g., the context information related to the sensors needed to be embedded into the data set, because real data that matches our context framework is not available in any public data sets at the moment) to better understand the behaviour of CASSARAM in different IoT related real world situations and scenarios where real data is not available. We make the following assumptions in our work. We assume that the sensor descriptions and context information related to the sensors have already been retrieved from the sensor manufacturers in terms of ontologies, and been into the SSN ontology. Similarly, we assume that the context data related to the sensors, such as accuracy, reliability, etc., have been continually monitored, measured, managed, and stored in the SSN ontology by the software systems. In order to evaluate the distributed processing techniques, we proposed  an experimental test involving four computational nodes. All the nodes are connected to a private organizational network (i.e., The Australian National University IT Network). The hardware configurations of the three additional devices are as follows: (1) Intel Core i7 CPU with 6GB RAM, (2) Intel Core i5 CPU with 4GB, and  (3)  Intel Core i7 with 4G. The details are presented in Section \ref{sec:E:Evaluating_Distributed_Sensor_Searching}.

\textcolor{blue}{We evaluated the performance of CASSARAM using different combinations of relational operators, such as $<, >, =, \leq, \geq$. The scenarios numbered in Figs. \ref{Figure:Results9_Relational_Expression_Single}--\ref{Figure:Results12_Relational_Expression_Double_Memory_Usage} correspond to the scenario numbers listed below. All the experiments retrieve five context properties. (1) Do not use any relational operator. (2) 1 out of 5 context properties are restricted by $\geq$ (e.g., the accuracy is to be greater than 80\%) (3) 2 out of 5 (e.g., the accuracy is to be greater than 80\% AND reliability greater than 85\%), (4) 3 out of 5, (5) 4 out of 5. All 5 context properties are restrained (6) by $\geq$, (7) by $\leq$, (8) by $=$, (9) by $<$, (10) by $>$. (11) 1 out of 5 context properties are restricted by two relational operators (e.g., the accuracy is to be greater than $\geq$ 80\% AND less than $\leq$ 95\%), (12) 2 out of 5, (13) 3 out of 5, (14) 4 out of 5; All 5 context properties are restrained (15) by $\leq$ and $\geq$, (16) by $<$ and $>$. We increased the number of restrictions imposed using additional relational operators. (17) defined two ranges for each context properties (e.g., (accuracy $\geq$ 80\% AND $\leq$ 95\%) OR (accuracy $\geq$ 50\% AND $\leq$ 60\%)). (18) defined three ranges.}

\vspace{-4pt}

\section{Evaluation and Discussion}
\label{sec:Discussion}

\begin{figure*}[t!]

        \centering
        \begin{subfigure}[b]{165pt}
                \centering
                \includegraphics[scale=0.38]{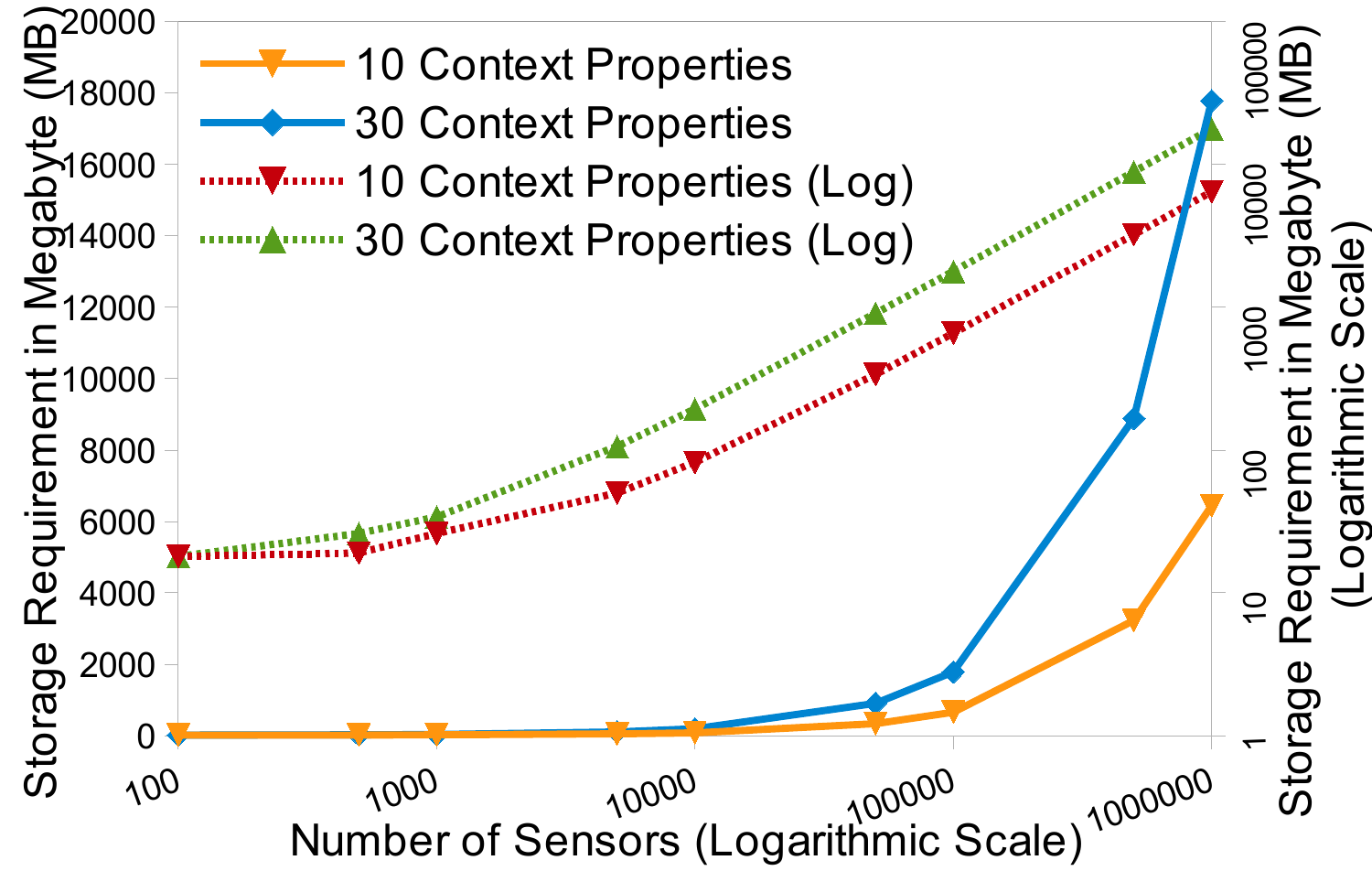}
                \vspace{-16pt}
                \caption{\footnotesize }
                \label{Figure:Results1_Storage_Requirment}
        \end{subfigure}%
        ~ %add desired spacing between images, e. g. ~, \quad, \qquad etc. 
        \begin{subfigure}[b]{165pt}
                \centering
                \includegraphics[scale=.38]{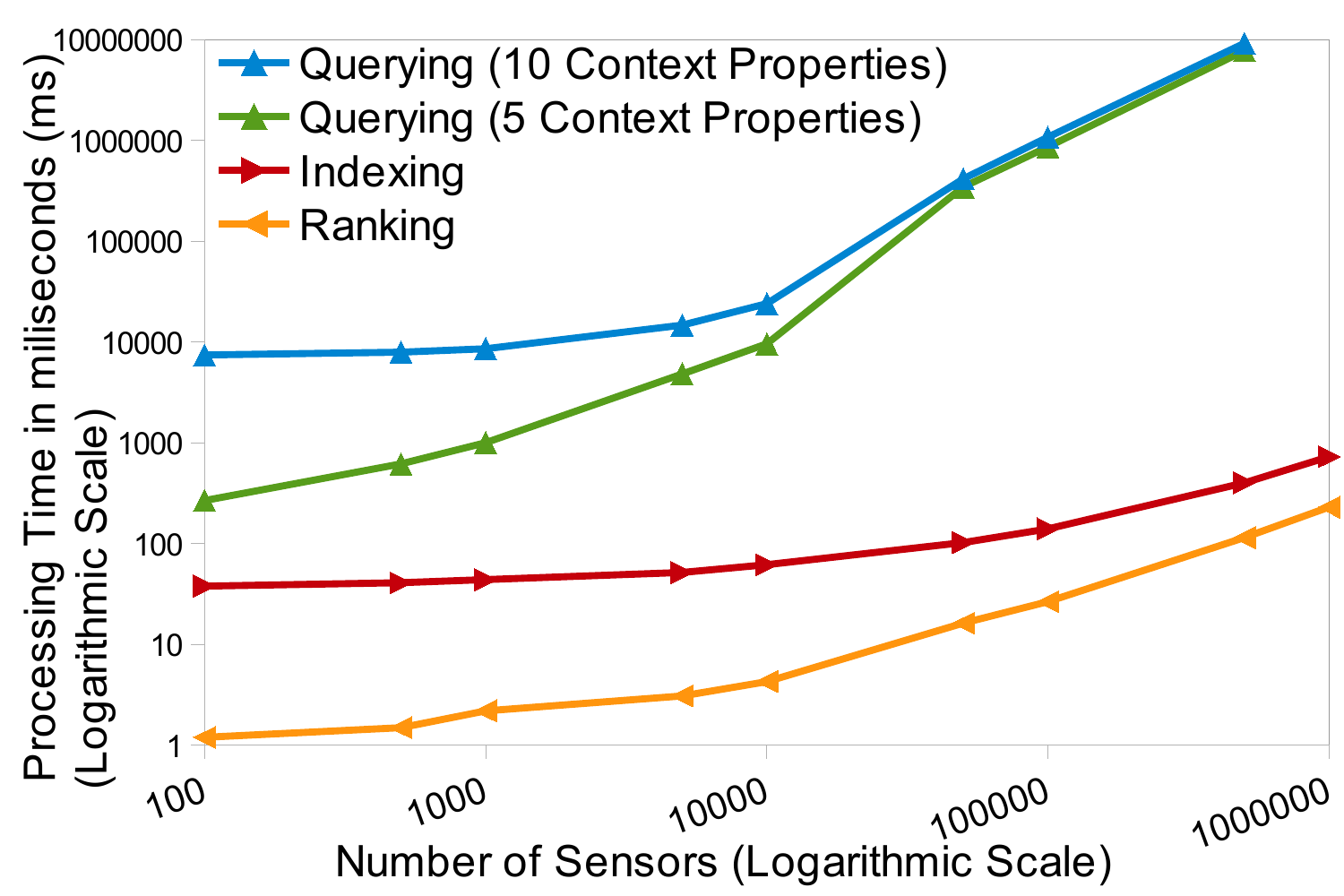}
                \vspace{-16pt}
                \caption{\footnotesize }
                \label{Figure:Results2_Overall_Processing_Time}
        \end{subfigure}
        ~ %add desired spacing between images, e. g. ~, \quad, \qquad etc. 
          %{0.3\textwidth}
        \begin{subfigure}[b]{165pt}
                \centering
                \includegraphics[scale=.38]{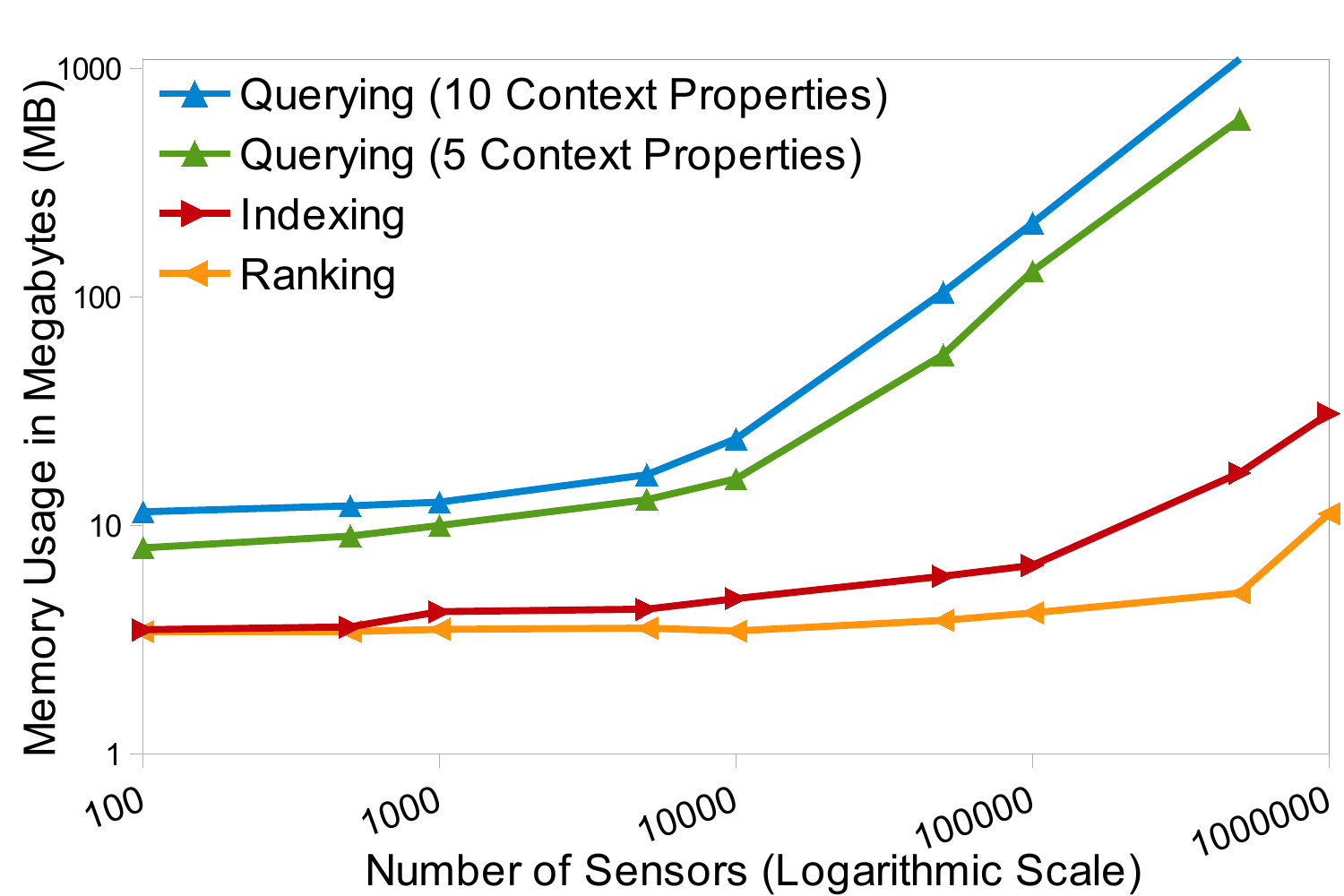}
                \vspace{-16pt}
                \caption{\footnotesize }
                \label{Figure:Results3_Overall_Memory_Usage}
        \end{subfigure}
        %\caption{Pictures of animals}\label{fig:animals}
\end{figure*}

\begin{figure*}
\vspace{-12pt}
        \centering
        \begin{subfigure}[b]{165pt}
                \centering
                \includegraphics[scale=0.38]{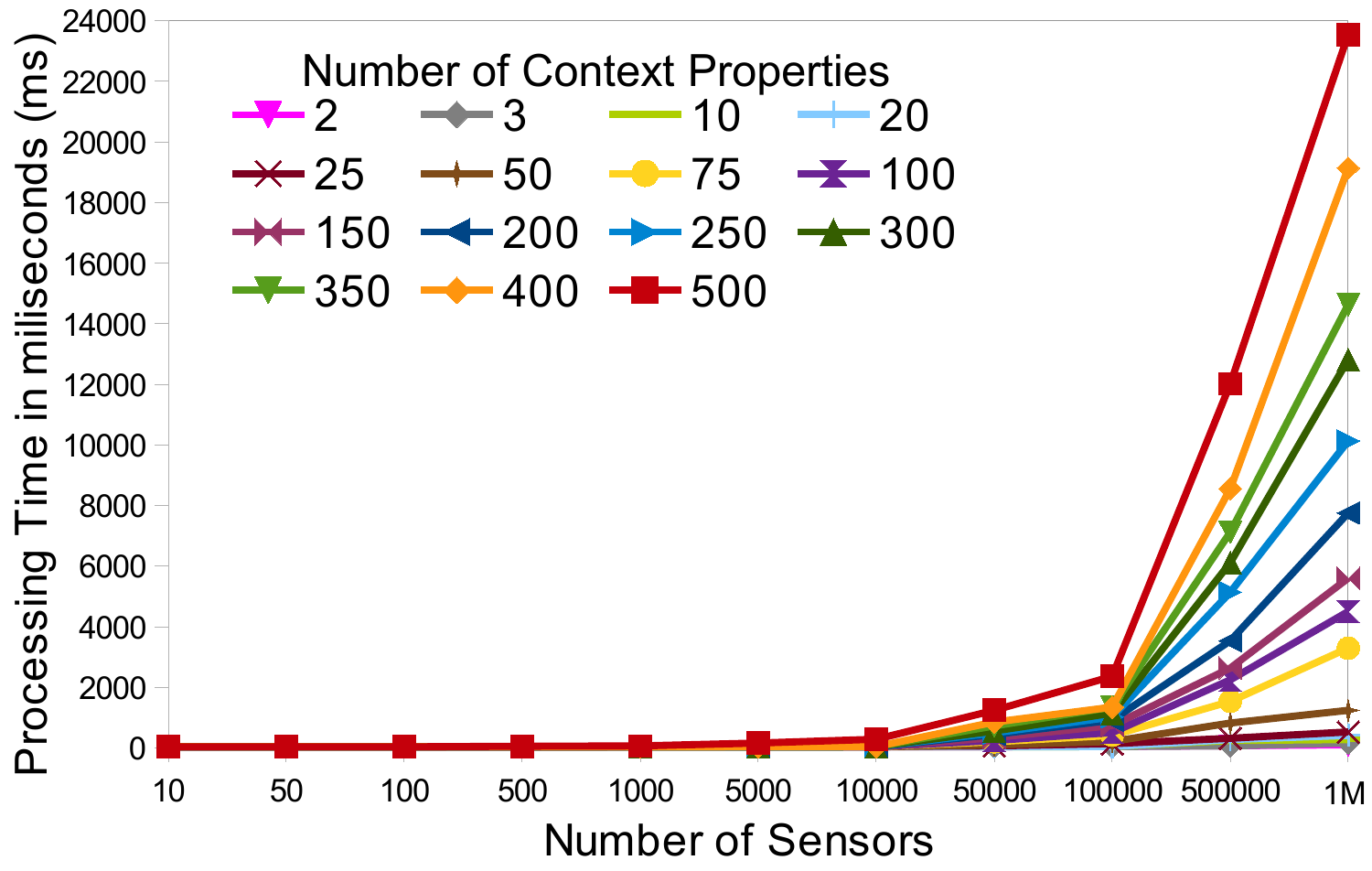}
                \vspace{-16pt}
                \caption{\footnotesize }
                \label{Figure:Results4_Context_Properties_Processing_Time}
        \end{subfigure}%
        ~ %add desired spacing between images, e. g. ~, \quad, \qquad etc. 
        \begin{subfigure}[b]{165pt}
                \centering
                \includegraphics[scale=.38]{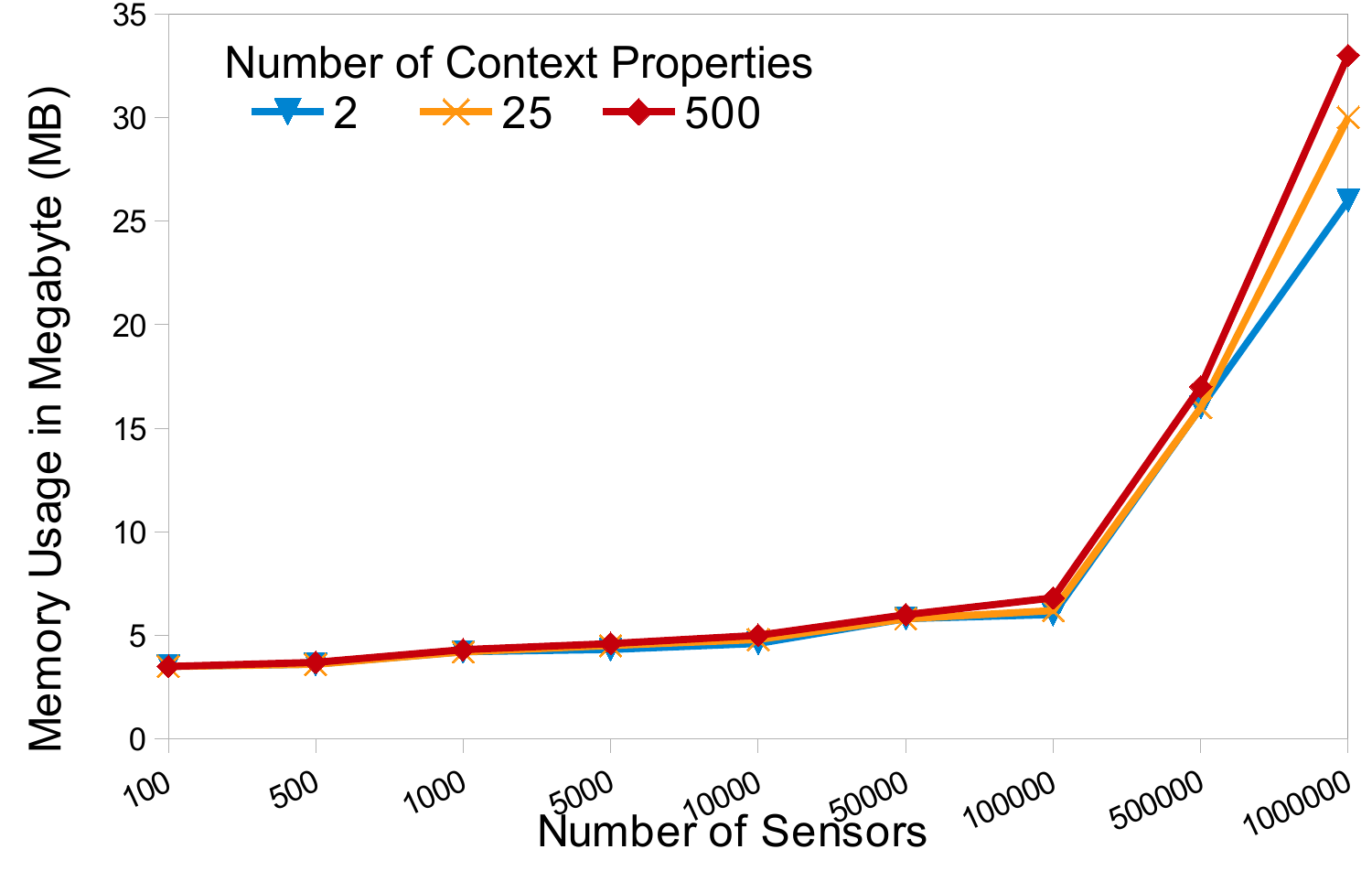}
                \vspace{-16pt}
                \caption{\footnotesize }
                \label{Figure:Results5_Context_Properties_Memory_Usage}
        \end{subfigure}
        ~ %add desired spacing between images, e. g. ~, \quad, \qquad etc. 
          %{0.3\textwidth}
        \begin{subfigure}[b]{165pt}
                \centering
                \includegraphics[scale=.38]{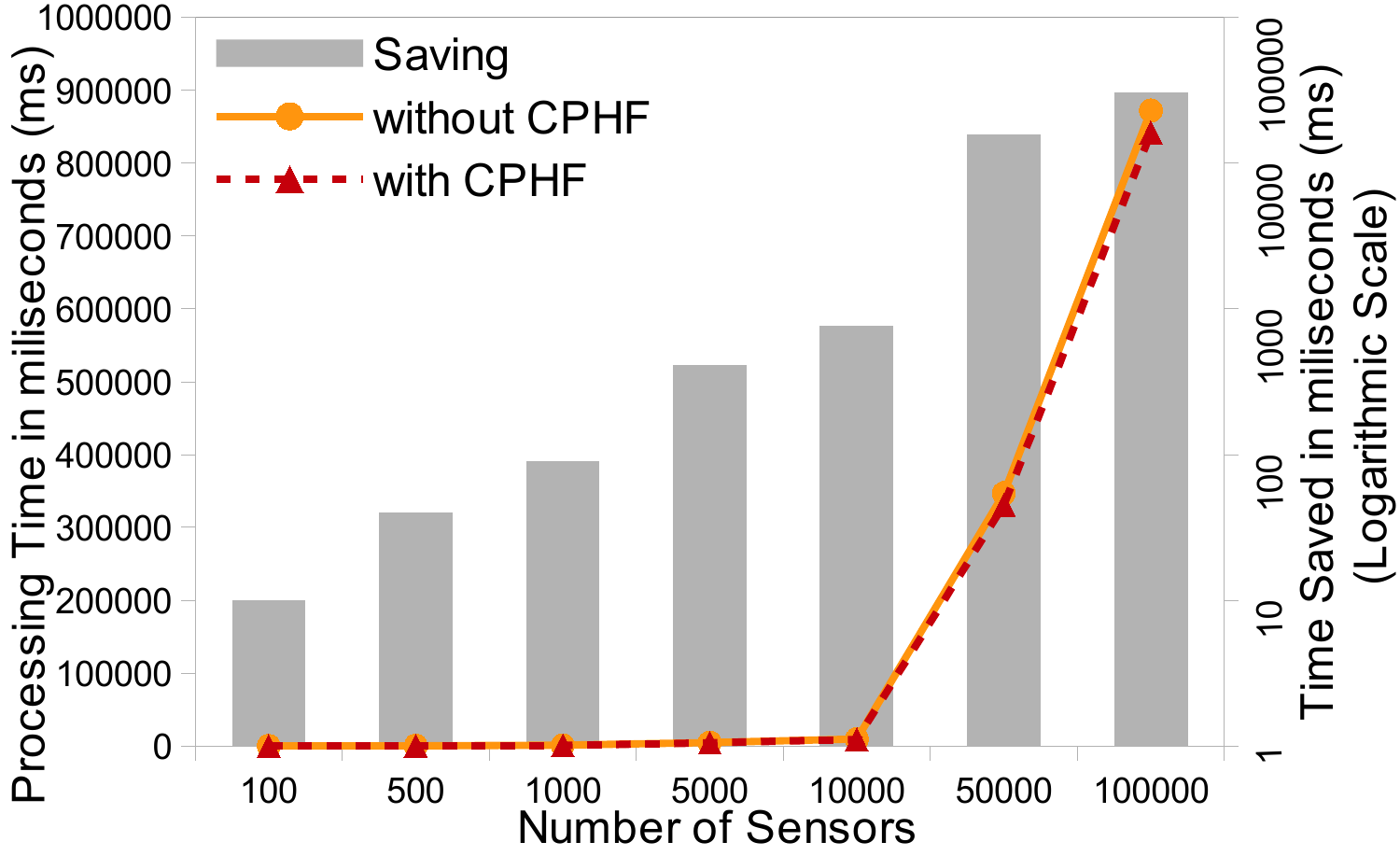}
                \vspace{-16pt}
                \caption{\footnotesize }
                \label{Figure:Results6_CPHF_Processing_Time}
        \end{subfigure}
        %\caption{Pictures of animals}\label{fig:animals}
         
\end{figure*}

\begin{figure*}
\vspace{-14pt}
        \centering
        \begin{subfigure}[b]{165pt}
                \centering
                \includegraphics[scale=0.38]{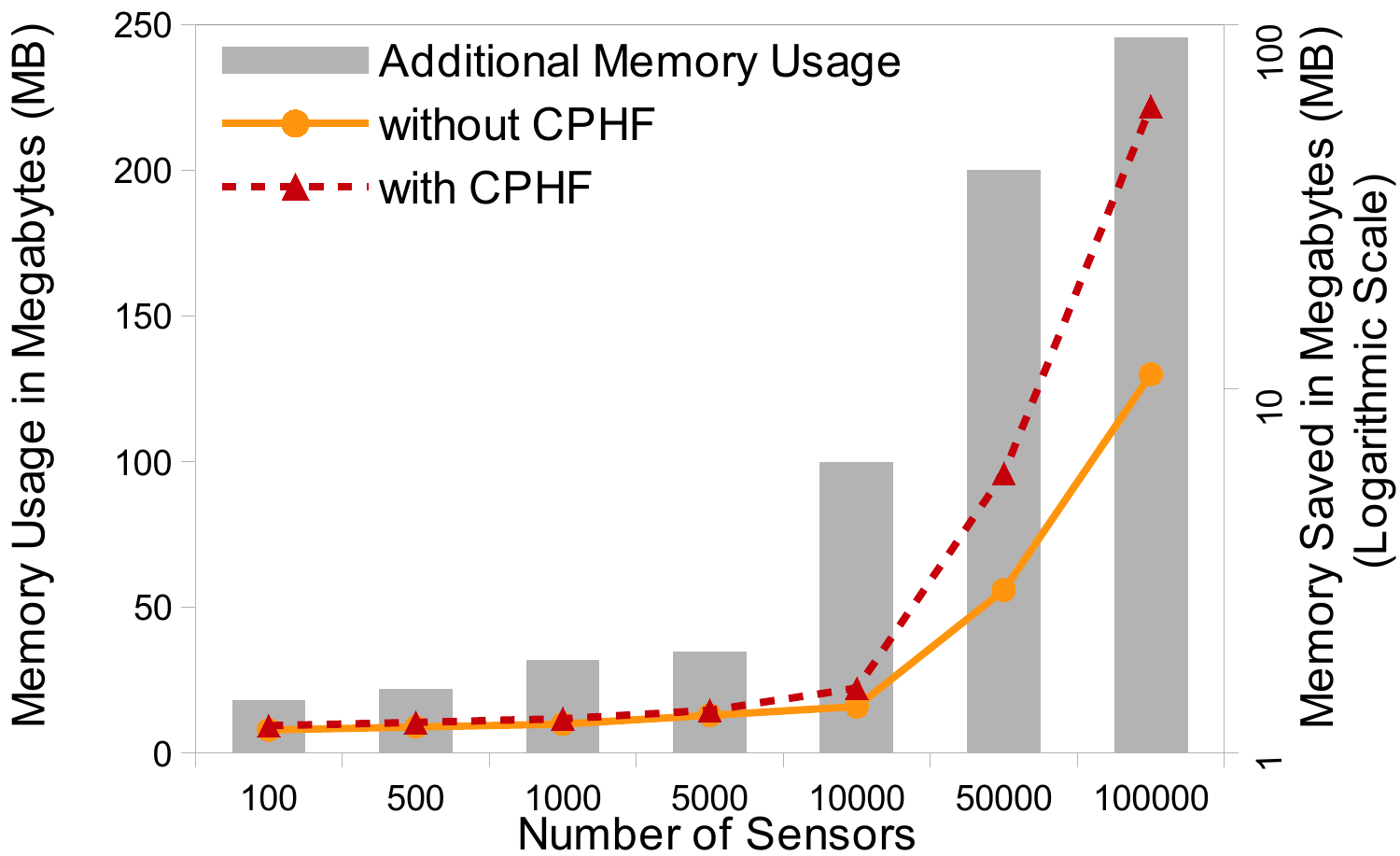}
                \vspace{-16pt}
                \caption{\footnotesize }
                \label{Figure:Results7_CPHF_Memory_Usage}
        \end{subfigure}%
        ~ %add desired spacing between images, e. g. ~, \quad, \qquad etc. 
        \begin{subfigure}[b]{165pt}
                \centering
                \includegraphics[scale=.38]{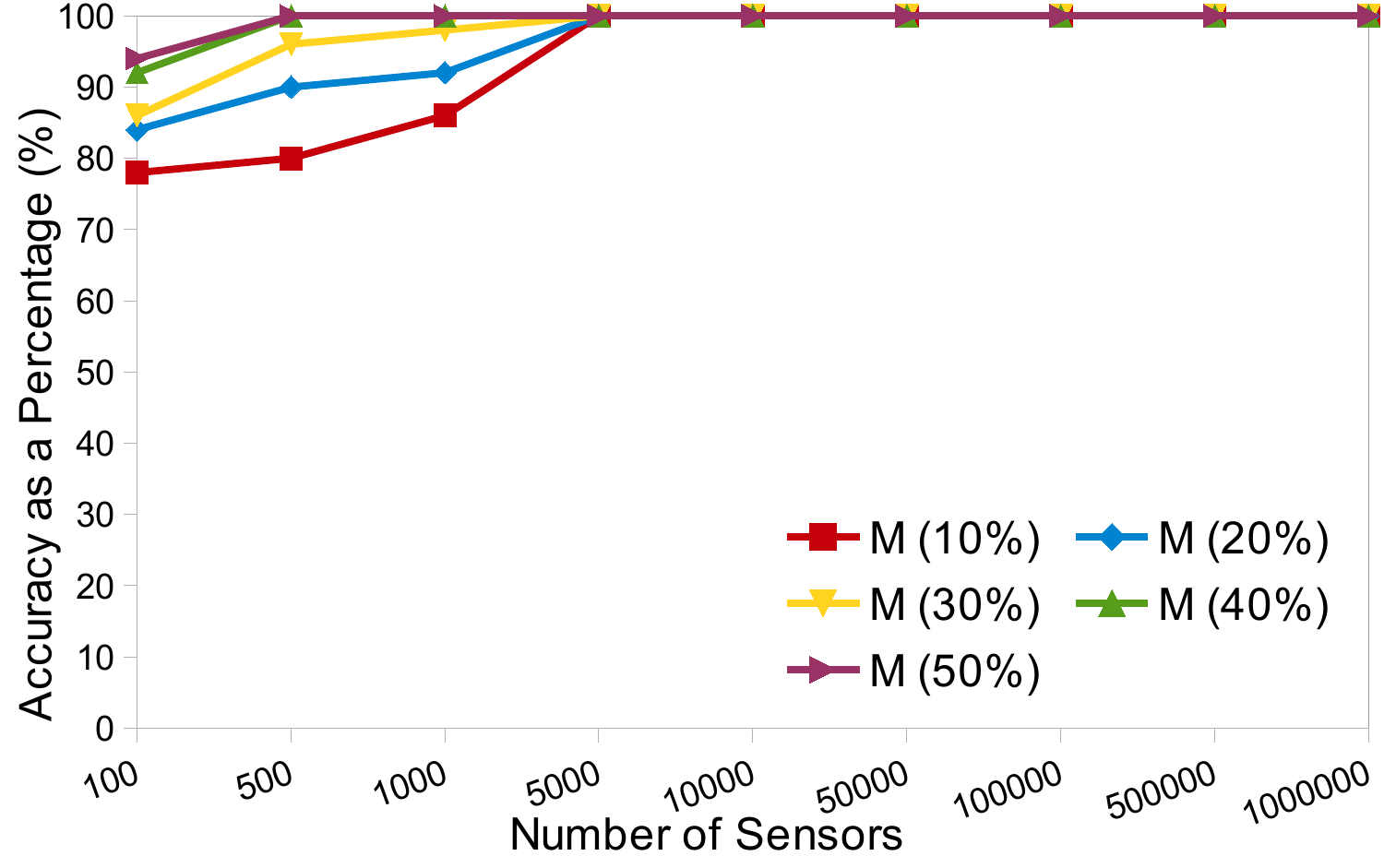}
                \vspace{-16pt}
                \caption{\footnotesize }
                \label{Figure:Results8_CPHF_Accuracy}
        \end{subfigure}
        ~ %add desired spacing between images, e. g. ~, \quad, \qquad etc. 
          %{0.3\textwidth}
        \begin{subfigure}[b]{165pt}
                \centering
                \includegraphics[scale=.38]{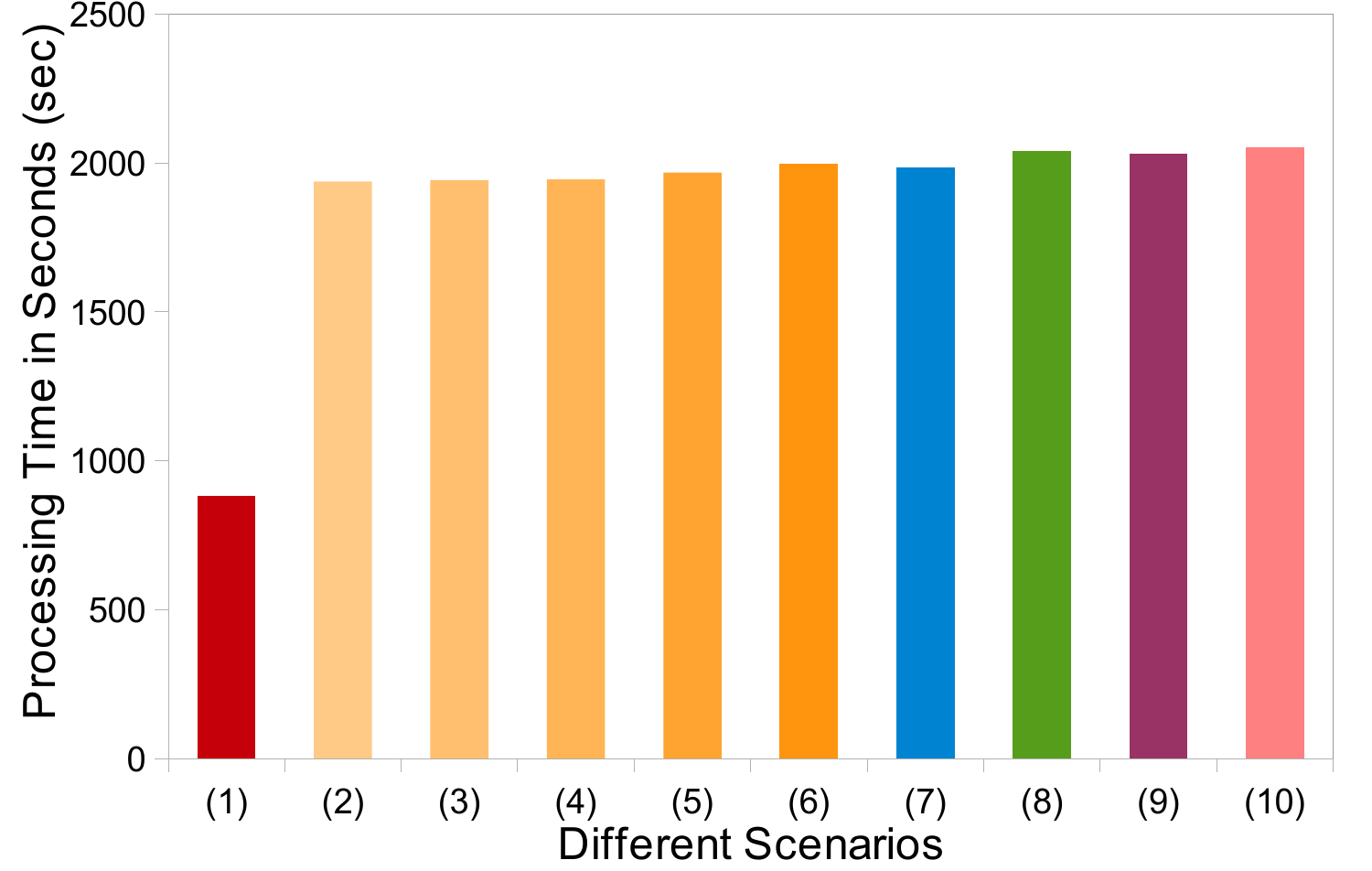}
                \vspace{-16pt}
                \caption{\footnotesize }
                \label{Figure:Results9_Relational_Expression_Single}
        \end{subfigure}
        %\caption{Pictures of animals}\label{fig:animals}
\end{figure*}

\begin{figure*}
\vspace{-12pt}
        \centering
        \begin{subfigure}[b]{165pt}
                \centering
                \includegraphics[scale=0.38]{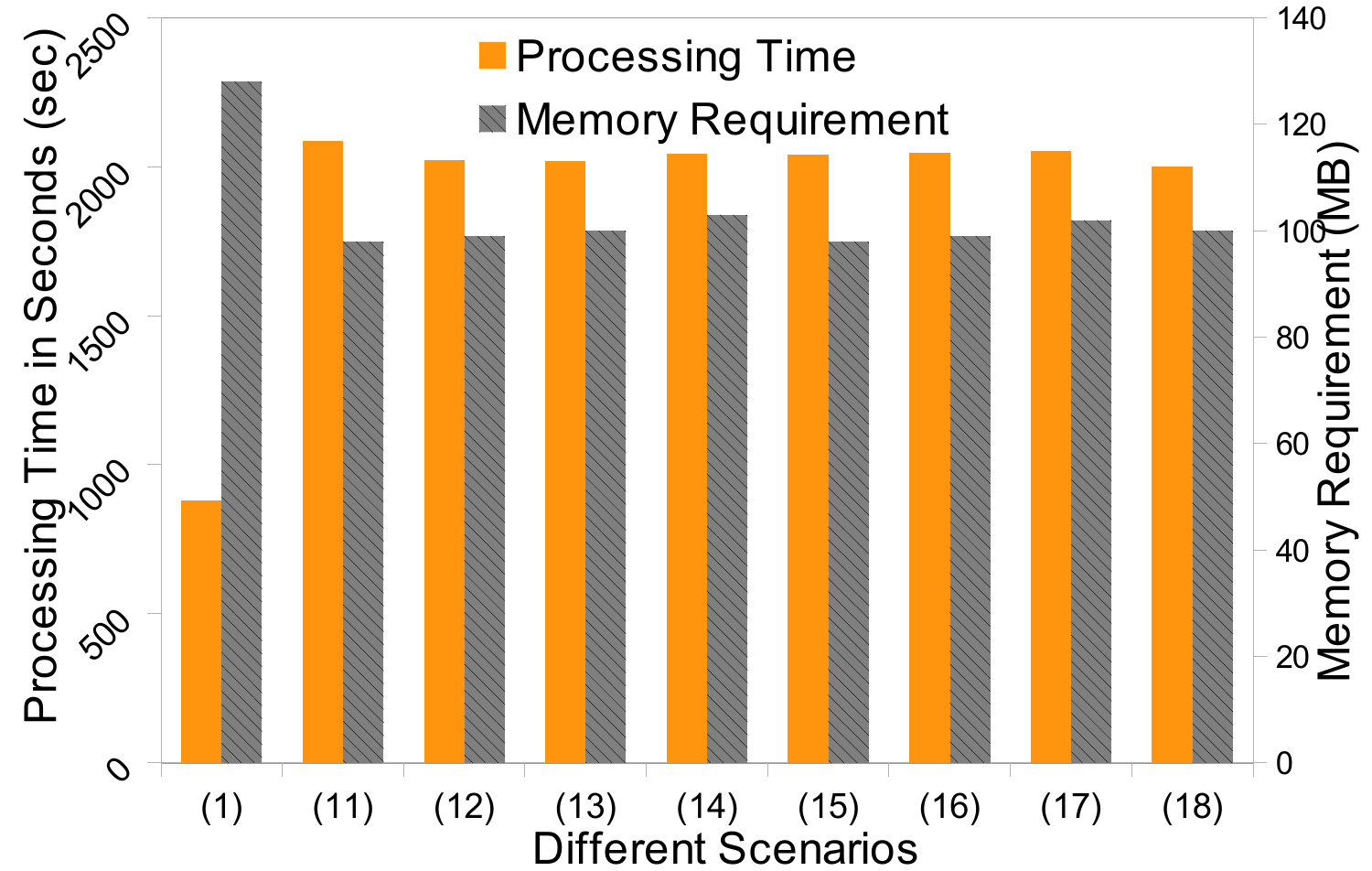}
                \vspace{-16pt}
                \caption{\footnotesize }
                \label{Figure:Results10_Relational_Expression_Double}
        \end{subfigure}%
        ~ %add desired spacing between images, e. g. ~, \quad, \qquad etc. 
        \begin{subfigure}[b]{165pt}
                \centering
                \includegraphics[scale=.38]{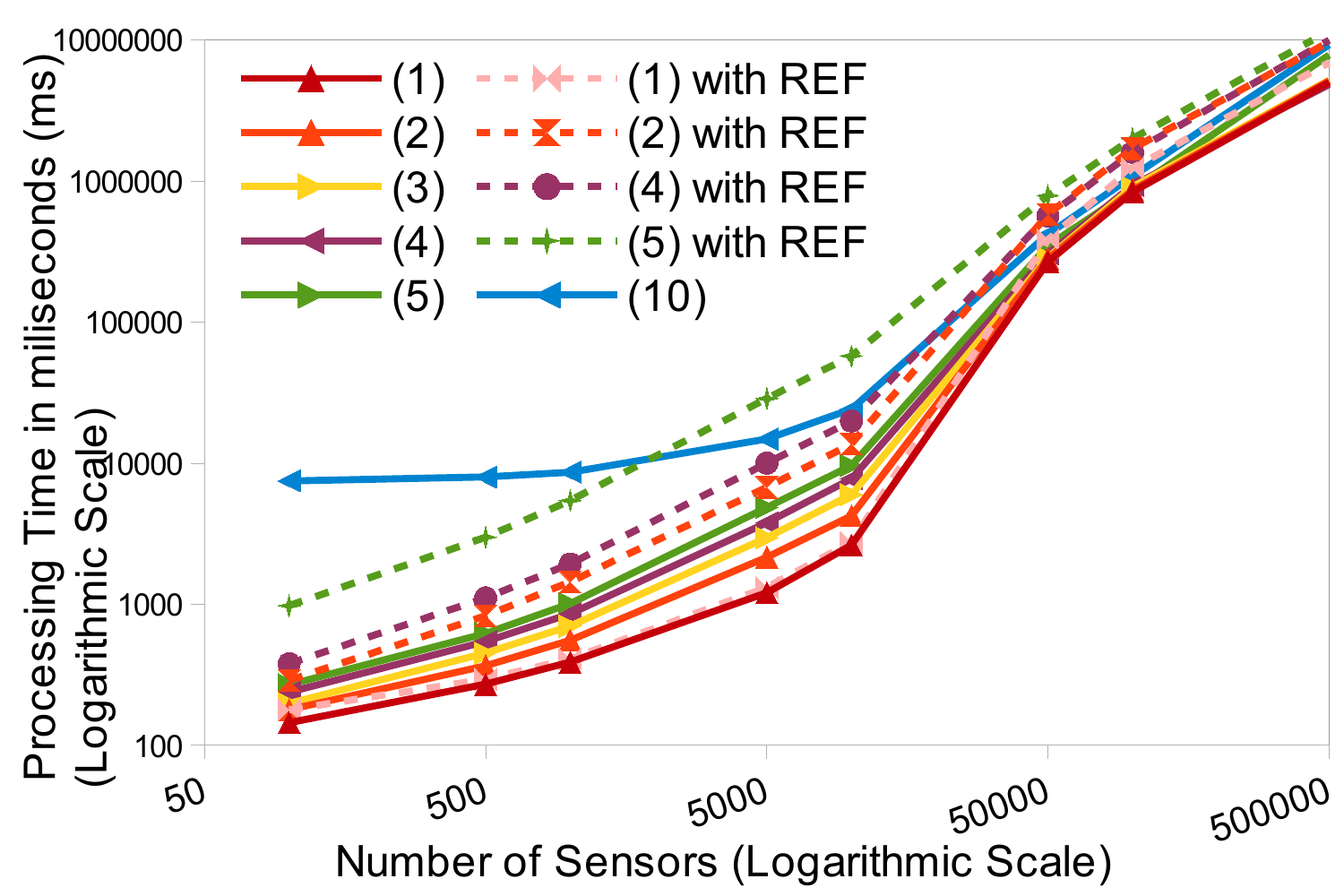}
                \vspace{-16pt}
                \caption{\footnotesize }
                \label{Figure:Results11_Relational_Expression_Double_10}
        \end{subfigure}
        ~ %add desired spacing between images, e. g. ~, \quad, \qquad etc. 
          %{0.3\textwidth}
        \begin{subfigure}[b]{165pt}
                \centering
                \includegraphics[scale=.38]{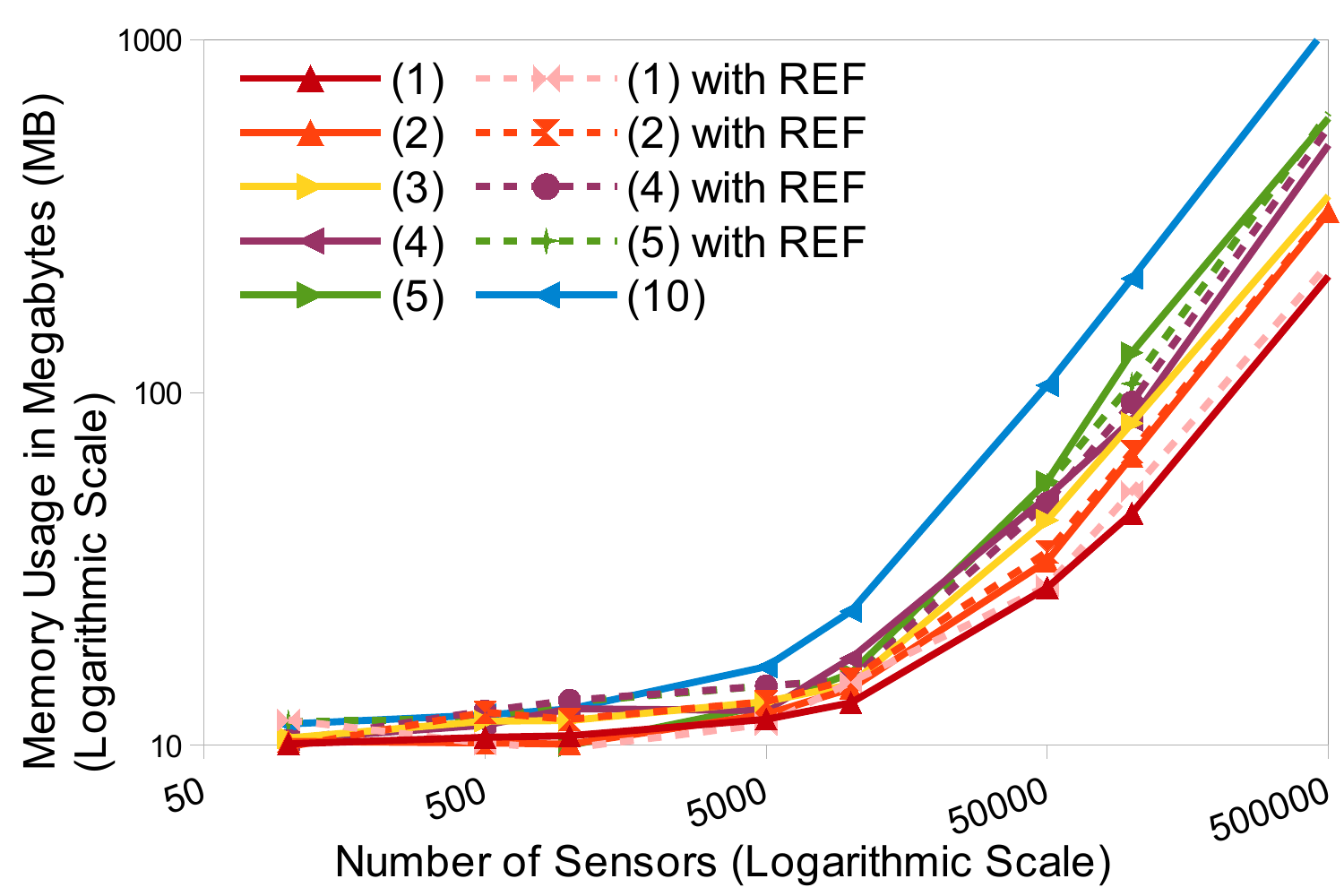}
                \vspace{-16pt}
                \caption{\footnotesize }
                \label{Figure:Results12_Relational_Expression_Double_Memory_Usage}
        \end{subfigure}
        \vspace{-4pt}
        \caption{Experimental Results}
        \label{Figure:Results}
\vspace{-20pt}
\end{figure*}

\textcolor{blue}{We evaluated CASSARAM using different methods and parameters as depicted in Figs. \ref{Figure:Results1_Storage_Requirment}--\ref{Figure:Results12_Relational_Expression_Double_Memory_Usage}. In this section, we explain the evaluation criteria which we used for each experiment and discuss the lessons we learned. Fig. \ref{Figure:Results1_Storage_Requirment} shows how the storage requirement varies depending on the number of sensor descriptions. We stored the data according to the SSN ontology, as depicted in Fig. \ref{Figure:Data_Model}. We conducted two experiments where we stored 10 context properties and 30 context properties from the context framework we proposed in Section \ref{sec:OA:Context-aware_QoS_Matrix}. To store one million sensor descriptions, it took 6.4 GB (10 context properties) and 17.8 GB (30 context properties). It is evident that the storage requirements are correlated with the number of triples: a single triple requires about 0.193 KB storage space (for 100,000+ sensors). Though storage hardware is becoming cheaper and available in high capacities, the number of context properties need to store should be decided carefully in order to minimize the storage requirements, especially when the number of sensor is in the billions.}

\textcolor{blue}{Fig. \ref{Figure:Results2_Overall_Processing_Time} shows how much time it takes to select sensors as the number of sensors increases. Each step (i.e., searching, indexing and  ranking) has been measured separately. Semantic querying requires significantly more processing time than indexing and  ranking. Furthermore, as the number of context properties retrieved by a query increases, the execution time  also increases significantly. Furthermore, it is important to note that MySQL can  join only 61 tables, which only allows retrieving a maximum of 10 context properties from the SSN ontology data model. Using alternative data storage or running multiple queries can be used to overcome this problem. Similarly, it is much more efficient to run multiple queries than to run a single query if the number of sensors is less than 10,000 (e.g., 8 ms to retrieve 5 context properties and 24 ms to retrieve 10 context properties when querying 10,000 sensors).} \textcolor{blue}{In addition, Fig. \ref{Figure:Results3_Overall_Memory_Usage} shows how much memory is required to select sensors as the number of sensors increases. It is evident that having more context properties requires having more memory. The memory requirements for querying do not change much up to 10,000 (ranging from 10 MB to 25 MB). When the number of sensors exceeds 10,000, the memory requirements grow steadily, correlated with the number of sensors. In comparison, indexing and ranking require less memory.}

\textcolor{blue}{Fig. \ref{Figure:Results4_Context_Properties_Processing_Time} shows the processing time taken by the sensor indexing process as the number of context properties and the number of sensors increase. Reducing the number of sensors needing to be indexed below 10,000 allows speeding up CASSARAM. The processing time starts to increases significantly after 100,000 sensors. Similarly, Fig. \ref{Figure:Results5_Context_Properties_Memory_Usage} shows the memory usage by the sensor indexing process as the number of context properties and sensors increases. Even though the memory requirements increase slightly, the actual increase is negligible when the number of sensors is still less than 100,000. After that, the memory requirements increase substantially, but are still very small compared to the computational capabilities of the latest hardware. Furthermore, the number of context properties involved does not have any considerable impact during the indexing process. The differences only  become visible when the number of sensors reaches one million. Still, the memory required by the process is 30 MB. Java garbage collection performs its task more actively when processing large numbers  of sensors, which makes the difference invisible.}

\textcolor{blue}{Fig. \ref{Figure:Results6_CPHF_Processing_Time} and \ref{Figure:Results7_CPHF_Memory_Usage} compare the time taken by the sensor selection process and the memory it requires, with and without the CPHF algorithm, as the number of sensors increases. The number of sensors that the user requires is kept at 50 in all experiments (${N}$=50). Five context properties are retrieved, indexed, and ranked. The complexity of CPHF (due to the SPARQL subqueries) has not affected significantly the total processing time of CASSARAM. Instead, CPHF has saved some time in the indexing and ranking phases. In contrast, CPHF requires more memory when querying, due to its complexity. However, it requires significantly less memory when transferring data to the next phase for indexing. Therefore, CPHF is efficient as it does not require holding millions of pieces of sensor information in multiple phases in CASSARAM. Furthermore, CPHF returns only a limited number of sensors whereas the non-CPHF approach returns all sensors available to CASSARAM, which consumes more resources including more processing time and a significant amount of memory and temporary storage.} \textcolor{blue}{Fig. \ref{Figure:Results8_CPHF_Accuracy}  shows how the accuracy changes when the Margin of Error (M\%) value changes in the CPHF algorithm and the number of sensors increases. The scenario presented in Fig. \ref{Figure:Comparative_Priority-based_Heuristic_Filtering} has been evaluated. The accuracy of the CPHF approach increases when the margin of error (\textit{M}) increases. However, a lower \textit{M} leads CASSRAM towards low resource consumption. Therefore, there is a trade-off between accuracy and resource consumption. The optimum value of \textit{M} can be dynamically learned by machine learning techniques based on which context properties are prioritized by the users in each situation and how the normalized weights are distributed between the context properties.}

 \textcolor{blue}{In Fig. \ref{Figure:Results9_Relational_Expression_Single} and Fig. \ref{Figure:Results10_Relational_Expression_Double}, we evaluated how processing time and memory requirements change when relational expressions are used during the semantic querying phase. We tested different scenarios with and without relational expressions (e.g. $<,>,=,\leq,\geq$) as described  at the end of Section \ref{sec:Implementation}. For all experiments, we queried 100,000 sensors. When at least one relation operator is used in SPARQL, the processing time and the memory requirements increase by 100\%. However, neither the number of relational operators used nor the type of relational operators used make any impact on either processing time or memory requirements. Therefore, it is efficient to use multiple relational operators (as much as possible) so as to reduce the number of sensors retrieved by the querying phase. This helps to reduce the amount of data needing to be handled in the other phases.}

\begin{figure*}[t]
%\vspace{-12pt}
        \centering
        \begin{subfigure}[b]{165pt}
                \centering
                \includegraphics[scale=0.38]{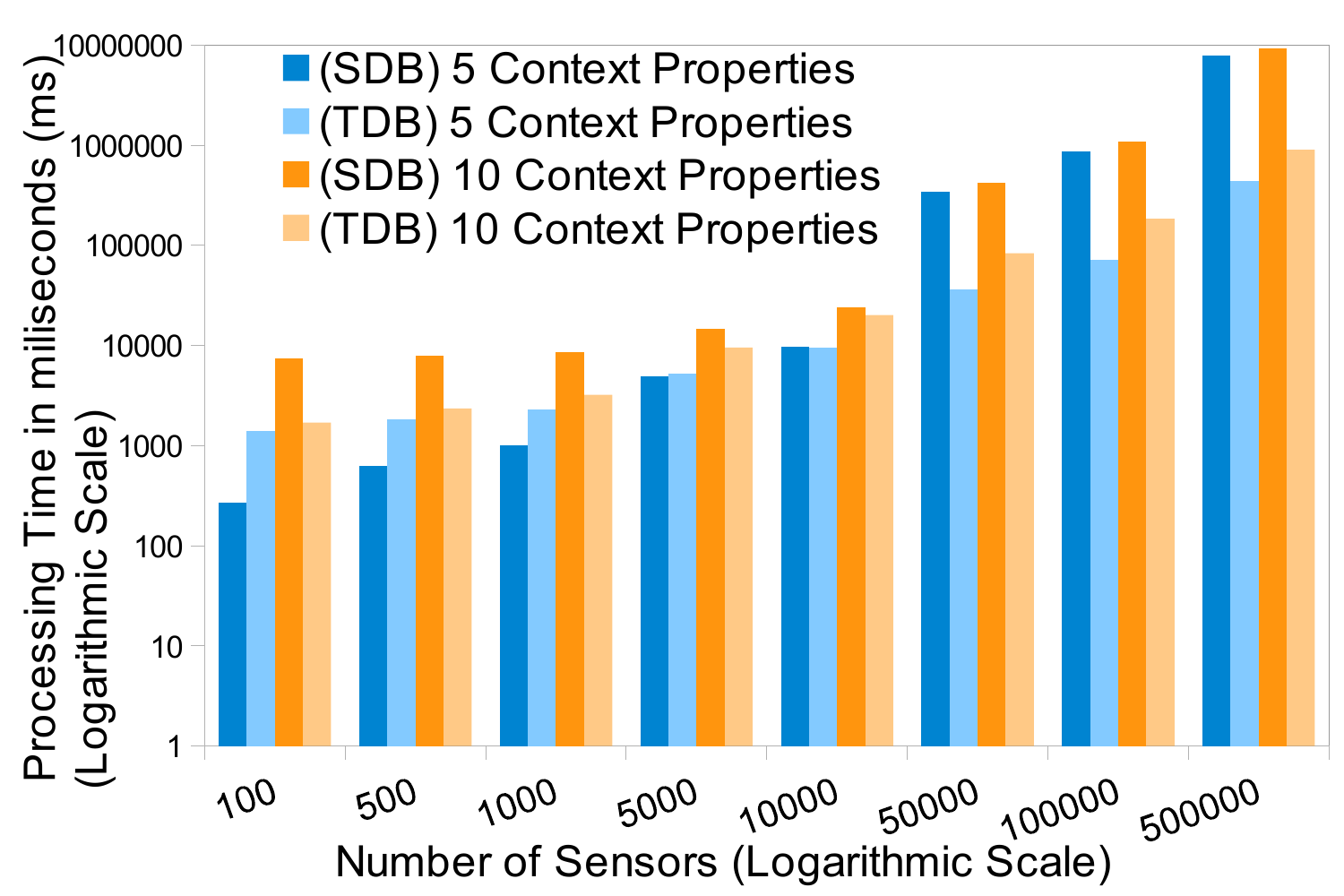}
                \vspace{-16pt}
                \caption{\footnotesize }
                \label{Figure:Results13_Data_Store_Comparison_Processing_Time}
        \end{subfigure}%
        ~ %add desired spacing between images, e. g. ~, \quad, \qquad etc. 
        \begin{subfigure}[b]{165pt}
                \centering
                \includegraphics[scale=.38]{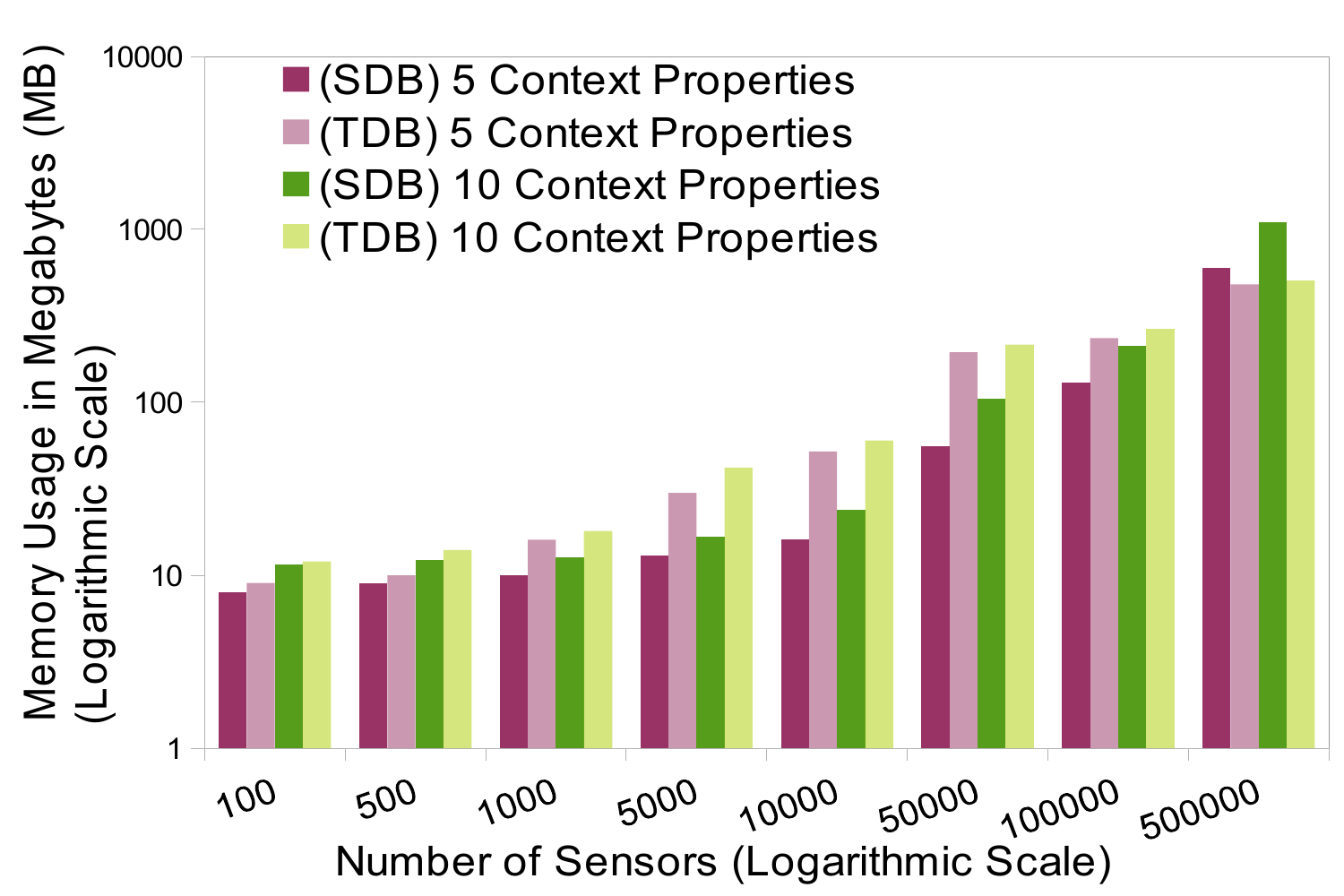}
                \vspace{-16pt}
                \caption{\footnotesize }
                \label{Figure:Results14_Data_Store_Comparison_Memory_Usage}
        \end{subfigure}
        ~ %add desired spacing between images, e. g. ~, \quad, \qquad etc. 
          %{0.3\textwidth}
        \begin{subfigure}[b]{165pt}
                \centering
                \includegraphics[scale=.38]{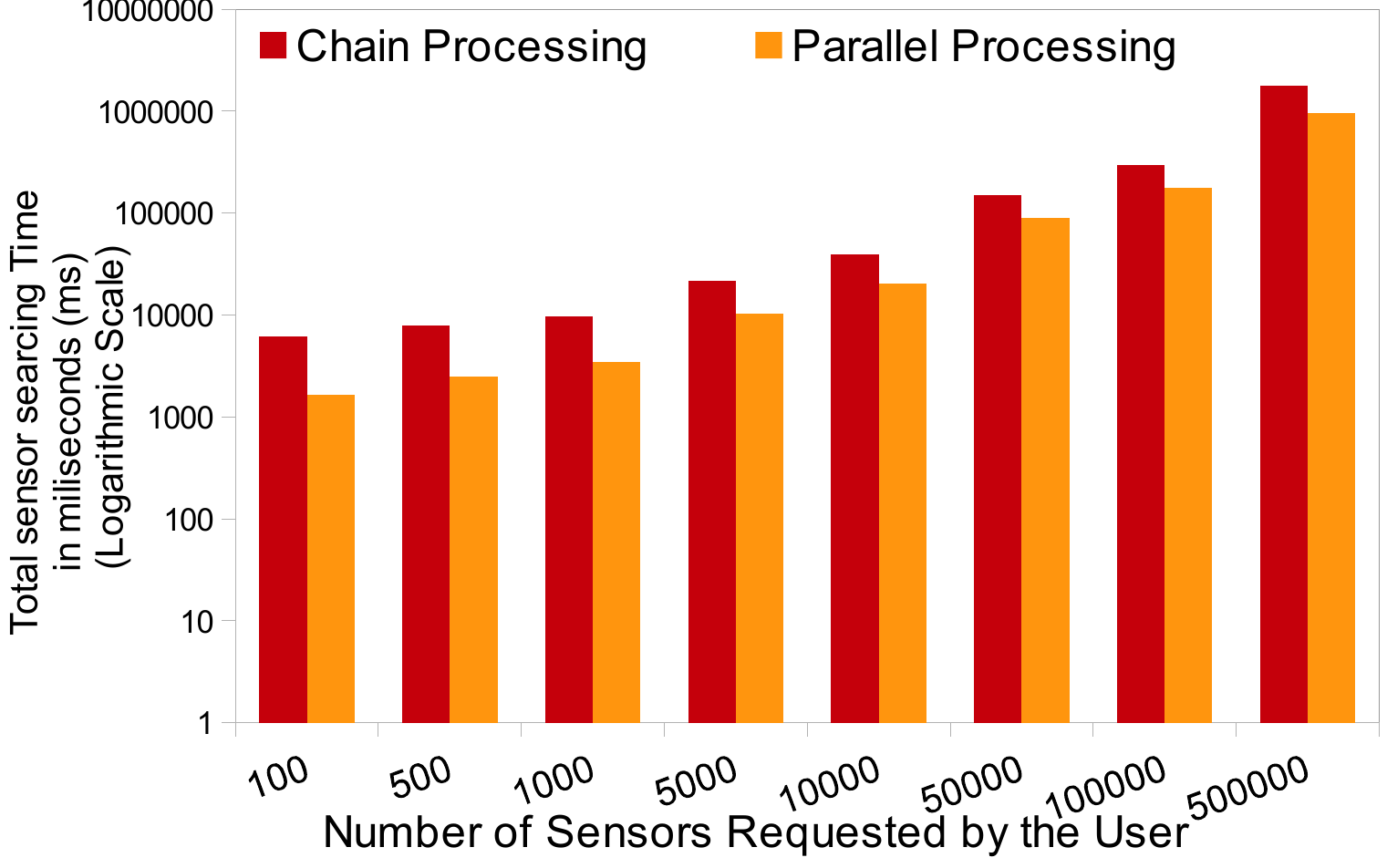}
                \vspace{-16pt}
                \caption{\footnotesize }
                \label{Figure:Results15_Distributed_Sensor_Searching}
        \end{subfigure}
        \vspace{-4pt}
        \caption{Results of alternative storage usage and distributed sensor searching}
        \label{Figure:Results2}
\vspace{-20pt}
\end{figure*}

\textcolor{blue}{Finally, in Figs. \ref{Figure:Results11_Relational_Expression_Double_10} and  \ref{Figure:Results12_Relational_Expression_Double_Memory_Usage}, we  extensively evaluated how REF affects the processing time and memory requirements in CASSARAM, as the number of sensors and context properties increases. As we mentioned earlier, REF adds more processing overhead, which affects the processing time and memory. There is a significant difference in processing time when the number of sensors needing to be queried is less than 100,000. However, when the number of sensors increases beyond 100,000, the difference becomes insignificant. In contrast, the differences in memory requirements are negligible when the number of sensors is less than 10,000: but it starts to become visible after that. Furthermore, the processing time increases significantly after 10,000 sensors}. We also learned that allocating more memory for CASSARAM can speed up the entire sensor selection process. 

In contrast, CASSARAM can  also be used under limited resources though it takes a much longer time to respond. According to the extensive evaluations we conducted, it is evident that CPHF and REF techniques can be used to improve the efficient of CASSARAM. Even though this paper is specifically focused on sensor selection in the IoT domain, the proposed model and the concepts we employed can be used in many other domains, such as web service selection. Furthermore, the results we obtained through these evaluations are also applicable to any other approach that employs an ontology model similar to the SSN ontology and requires a large number of records. Even though we tested our solution with millions of sensor descriptions, in practice it is highly unlikely that millions of sensors would connect to a single middleware instance. Practically, IoT middleware solutions will store data in a distributed manner in different instances, and need to be searched in a distributed fashion, as explained in Section \ref{sec:CASSARAM:Context-aware_Distributed_Searching}. By parallel processing, the amount of time it takes to process millions of sensor data descriptions can be reduced drastically.

 %%%%%%%%%%%%%%%%%%%%%%%%%%%%%%%%%%%
\begin{table*}[b]
\centering
\footnotesize
\renewcommand{\arraystretch}{1.15}

\caption{The amount of redundant data communication saved by the parallel sensor search with $k$-extension strategy}
\vspace{-0.2cm}
\begin{tabular}{c m{1.5cm} c c c c c c c c c c c c  }
\hline
      \multicolumn{12}{c}{Number of sensors requested by the users ($N$)}\\
&     &100 &500 &1,000  &5,000  &10,000  &50,000    &100,000 & 500,000  &1,000,000 & \\ \hline \hline

\multirow{8}{*}{\rotatebox[origin=<O>]{90}{k value}} 
 & 10        & -60.7\llr  & -60.5\llr  & -60.3\llr  & -58.7\llr  & -56.7\llr  & -40.5\llr  & -20.2\llr  & 141.6\llg    & 344.0\llg  & \multirow{8}{*}{\rotatebox[origin=<O>]{90}{in Megabytes (MB)}}\\ 
 & 100             & \gc        & -5.9\lly   & -5.7\lly   & -4.1\lly   & -2.1\lly   & 14.1\llg   & 34.3\llg   & 196.2\llg    & 398.5\llg  & \\ 
 & 500             & \gc        & \gc  & -1.1\lly   & 0.5\llg    & 2.5\llg    & 18.7\llg   & 38.9\llg   & 200.8\llg    & 403.1\llg  & \\ 
 & 1000            & \gc        & \gc        & \gc        & 0.8\llg    & 2.8\llg    & 19.0\llg   & 39.3\llg   & 201.1\llg    & 403.5\llg  & \\ 
 & 5000            & \gc        & \gc        & \gc        & \gc        & 0.9\llg    & 17.1\llg   & 37.3\llg   & 199.2\llg    & 401.5\llg  & \\ 
 & 10000           & \gc        & \gc        & \gc        & \gc        & \gc        & 14.1\llg   & 34.3\llg   & 196.2\llg    & 398.5\llg  & \\ 
 & 50000           & \gc        & \gc        & \gc        & \gc        & \gc        & \gc        & 10.1\llg   & 172.0\llg    & 374.3\llg  & \\ 
 & 100000          & \gc        &  \gc       & \gc        & \gc        & \gc        & \gc        & \gc        & 141.6 \llg   & 344.0\llg  & \\ 
 & 500000          & \gc        & \gc        & \gc        & \gc        & \gc        &\gc         & \gc        & \gc          & 101.2\llg  & \\ 

\hline
\end{tabular}
\label{Tbl:Summarized Bandwidth_Saving}
%\vspace{-0.6cm}
\end{table*}
%%%%%%%%%%%%%%%%%%%%%%%%%%%%%%%%%%%

\vspace{-4pt}

\subsection{Evaluating Alternative Storage Options}
\label{sec:E:Evaluating_Alternative_Storage_Options}

In the evaluations conducted earlier (Figs. \ref{Figure:Results1_Storage_Requirment}--\ref{Figure:Results12_Relational_Expression_Double_Memory_Usage}), we used Jena SDB/MySQL-backed RDF storage to store the data. In order to evaluate the performance of CASSARAM when using alternative storage options, we here employ a Jena TDB-backed approach (jena.apache.org/documentation/tdb). In Fig. \ref{Figure:Results13_Data_Store_Comparison_Processing_Time}, we compare the processing times taken by both the Jena SDB/MySQL and the Jena TDB approach. Furthermore, in Fig. \ref{Figure:Results14_Data_Store_Comparison_Memory_Usage}, we compare the memory usage  by  the SDB and TDB approaches. According to the Berlin SPARQL Benchmark \cite{P591}, Jena TDB is much faster than Jena SDB. We also observed similar results both in 5 context data processing as well as in 10 context data processing. Specifically, Jena TDB is 10 times faster than SDB when processing 10 context properties, where the dataset consists of half a million sensor descriptions. The Jena SDB approach consumed less memory than the Jena TDB approach when the dataset was less than 100,000 sensor descriptions. However, after that, the Jena TDB approach consumes less memory than the Jena SDB. Specifically, Jena TDB uses 50\% less memory than Jena SDB when   processing 10 context properties, where the dataset consists of half a million sensor descriptions. Therefore it is evident that Jena TDB is more suitable when the number of sensor descriptions goes beyond 100,000.

Despite the differences we observed in our evaluation, there are several factors that need to be considered when selecting underlying storage solutions. As evaluated on the Berlin SPARQL Benchmark, there are several other storage options available, such as Sesame (openrdf.org), Virtuoso TS, Virtuoso RV, and D2R Server \cite{P591}. Jena TDB offers faster load times and better scale, but has the worst query performance. Sesame seems better all-round for low data sizes assuming infrequent loads. In contrast, Jena SDB provides moderate performance, offering load times, query performance, and scalability between the Jena TDB and Sesame. Based on these evaluations, at the time at which this paper was written, there is no superior solution that has all good qualities. Due to the lack of extensive usage and the short existence  of  Sesame, SDB/MySQL can be seen as a better choice especially when considering database functionalities such as backup, concurrent and parallel processing. As we do not expect frequent loading/ unloading of datasets such as sensor descriptions, it is evident that SDB  outperforms TDB in query processing (excluding data loading) \cite{P591}. As we expect more updates (transactions) to occur, SDB would be a better choice.

\subsection{Evaluating Distributed Sensor Searching}
\label{sec:E:Evaluating_Distributed_Sensor_Searching}

We evaluated distributed sensor searching using a private network that consists of four computational nodes. We compare two different distributed sensor search techniques, namely, chain processing and parallel processing with/without $k$-extensions, which we discussed in Section \ref{sec:CASSARAM:Context-aware_Distributed_Searching}. The results are presented in Fig. \ref{Figure:Results15_Distributed_Sensor_Searching}. Each node consists of a dataset of one million sensor data descriptions. The four datasets are different from each other. Five context properties are considered for the evaluation and the context information is stored using Jena TDB. First, we discuss the techniques from the theoretical perspective.

Let us define some of the notations which will be used in the following discussion: $n$= number of computational nodes (in our experiments $n$=4), $N$=number of sensors requested by the users, $S_{i}$= number of sensor descriptions stored in the $i$th computational node, $r$= size of a single sensor description record (i.e., storage requirements),  $t^{net}_{i,j}$= time taken for network communication between the computational nodes $i$ and $j$, $t^{pro}_{i}$= time taken to query the computational node $i$, merge the indexed results with the incoming results, and select the final number $N$ of sensors. The total time taken by chain-based distributed sensor searching can be defined as:

\vspace{-8pt}
\begin{equation}
%\vspace{-4pt}
Total_{chain} = \sum_{i=1}^{n} t^{pro}_{i} + \sum_{i=1}^{n-1} t^{net}_{i,i+1} + t^{net}_{n,1}
%\vspace{-8pt}
\end{equation}
\vspace{-8pt}

The total time taken by parallel distributed sensor searching can be defined as:
\vspace{-6pt}
\begin{equation}
%\vspace{-4pt}
Total_{parallel} = max  \left \{i=[2..n] : t^{pro}_{i} + t^{net}_{1,i}\right \}
%\vspace{-8pt}
\end{equation}

According to the results, it is evident that parallel processing is more efficient than chain processing in terms of the total processing time. However, parallel processing is inefficient in other aspects, such as network communication and bandwidth consumption. Therefore, we proposed $k$-extension to address this issue. The evaluation of the $k$-extension approach is presented in Table \ref{Tbl:Summarized Bandwidth_Saving}. In this experiment, we measured how much data communication can be saved (i.e., due to elimination of redundant data communication that occurs in parallel processing without $k$-extension) by using different $k$ values under different $N$ values. We measured the \textit{guaranteed minimum}\footnote{Depending on the dataset and the context information stored in each node, the parallel processing technique with $k$-extension will be able to save more data communication than the guaranteed minimum level.}  amount of data communications (measured in Megabytes) that can be saved.

In Table \ref{Tbl:Summarized Bandwidth_Saving}, positive values (marked in green) indicate the minimum amount of data communication saved using the $k$-extension. Although negative values (marked in orange/red) indicate no guaranteed savings, some situations (marked in orange) have a high chance of saving redundant data communication  compared to others. Equation (3) can be used to calculate the guaranteed minimum amount of data saving by using $k$- extensions. 
\vspace{-6pt}
\begin{multline}
Total_{Saving} = \sum_{i=2}^{n} S_{i}r - \left \{ [\sum_{i=2}^{n} \frac{S_{i}}{k} + N + (k-1)n] \times r  \right \} \\ \textrm{ IF } (k < N)
\end{multline}

Let us consider different scenarios where chain and parallel processing can be used. Chain processing is suitable for situations where saving computational resources and bandwidth is more critical than response time. A parallel processing method \textit{without $k$-extension} is suitable  when response time is critical and ${N}$ is fairly small. \textit{$k$-extension} requires two communication rounds: communication radios need to be opened and closed twice. Such a communication pattern consumes more energy \cite{ZMP001}, especially if the computational devices are energy constrained. Therefore,  transmitting data at once is more efficient. However, this recommendation becomes invalid when ${N}$ becomes very large (10,000+). Our experiments clearly show that $k$-extensions can be used to improve the efficiency of the parallel sensor searching approach, especially when ${N}$ is large. The ideal value of $k$ needs to be determined  based on $N$, $n$, and $S_{i}$.

\subsection{Application}
\label{sec:E:Application}

\begin{figure}[t]
\centering
%\vspace{-0.53cm}	
\includegraphics[scale=.17]{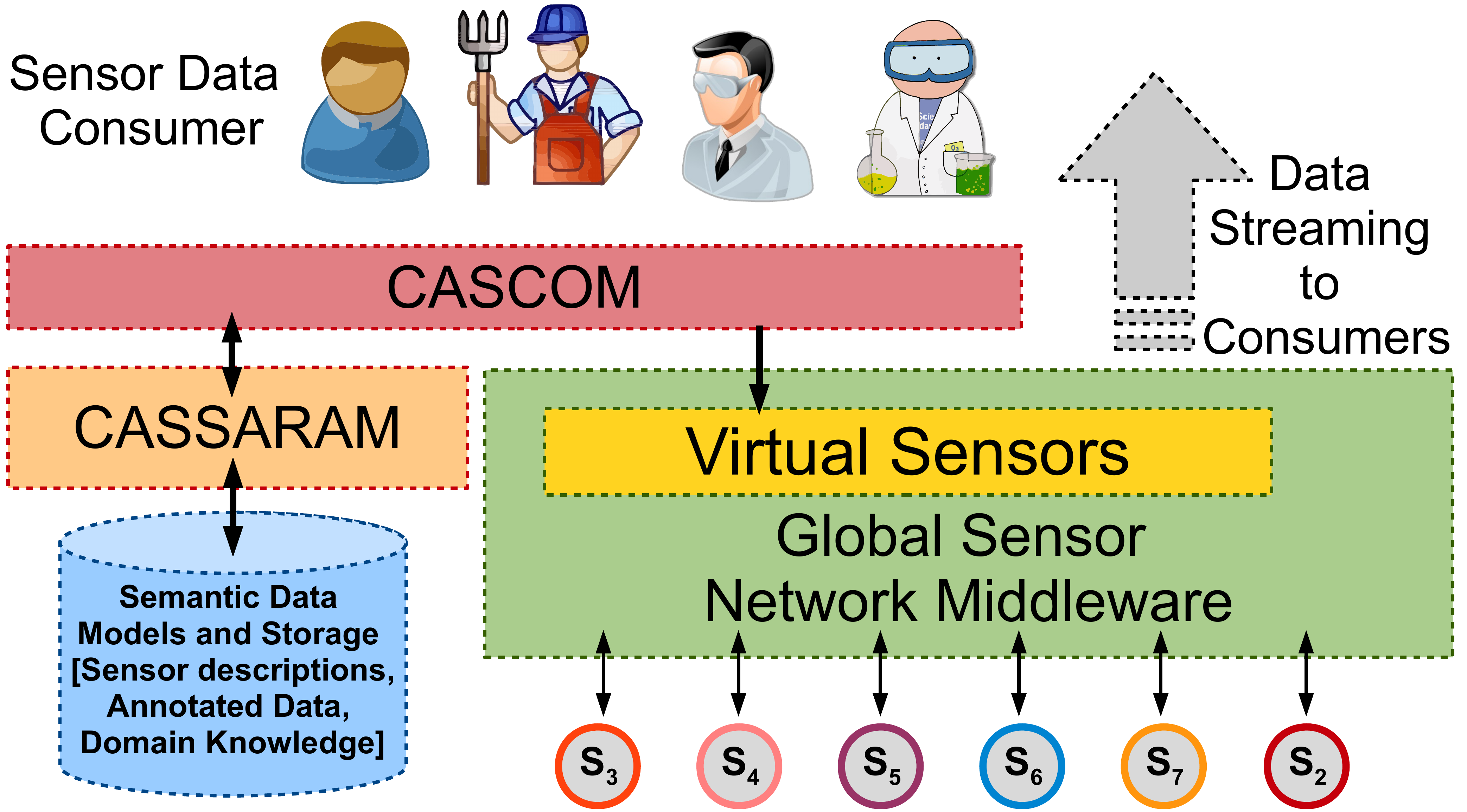}
\vspace{-0.18cm}	
\caption{CASSARAM in Action}
\label{Figure:Application}	
\vspace{-0.43cm}	
\end{figure}

In this section, we show where CASSARAM fits in the big picture (Figure \ref{Figure:Application}). Sensor data consumers are expected to interact with a model called\textit{ Context Aware
Sensor Configuration Model} (CASCoM) \cite{ZMP009}. Details explanation of CASCoM is out of the scope of this paper. Consumers are facilitated with a graphical user interface, which is based on a question-answer (QA) approach, that allows to express the requirements. Users can answer as many questions as possible. CASCoM searches and filters the tasks that the user may wants to perform. From the filtered list, users can select the desired task (e.g. environmental pollution detection). CASCoM searches for different programming components that allow to capture the data stream required by consumers (i.e. sensor data required to detect environmental pollution). CASCoM tries to find sensors that can be used to produce the inputs required by the selected data processing components. To achieve this task, CASCoM employs CASSARAM. Once the required sensor types are identified (and if multiple sensors are available), CASSARM graphical user interface is provided to the consumers to define their priorities. Later, the final set of sensors and data processing components are composed together. Required wrappers \cite{ZMP002} and the virtual sensor \cite{P022} are generated and sent to GSN by CASCoM. Finally, GSN starts streaming data to the consumer as defined in the virtual sensor.

%---------------- Section 9------------------%
\section{Conclusions and Future Research}
\label{sec:Conclusions and Future Work}

With advances in sensor hardware technology and cheap materials, sensors are expected to be attached to all the objects around us, which will increase the number of sensors available to be used. This means we have access to multiple sensors that would measure a similar environmental phenomenon. Such circumstances force us to choose between alternatives. We need to  decide which operational and conceptual sensor-related context properties are more important than others.

In this paper, we showed how the context information related to each sensor can be used to search and select the sensors that are best suited to a user's requirements. We  selected sensors based on  the user's expectations and priorities. As a proof of concept, we built a working prototype to demonstrate the functionality of our CASSARAM and to support the  experimentations using realistic applications. We also highlight how CASSARAM helps achieve our broader sensing-as-a-service vision in the IoT paradigm. CASSARAM allows optimizing the sensor data collection approaches by selecting the sensors in an optimized fashion. For example, CASSARAM can be used to find out which sensors have more energy and collect data only from those sensors. This helps to run the entire sensor network for a much longer time without reconfiguring. \textcolor{black}{We explored three different techniques that improve the efficiency and scalability of CASSARAM: comparative-priority based heuristic filtering, relational-expression based filtering, and distributed sensor searching.} We evaluated the performance of the proposed model extensively. In the future, we plan to incorporate CASSARAM  into leading IoT middleware solutions such as GSN, SenseMA, and OpenIoT, to support an automated sensor selection functionality in distributed environments. We will also investigate how to improve the efficiency of CASSARAM using \textcolor{black}{cluster-based sensor search} and heuristic algorithms that incorporate machine learning techniques.

\section{ACKNOWLEDGEMENT}
\label{sec:ACKNOWLEDGEMENT}

The authors acknowledge support from SSN TCP, CSIRO, Australia and ICT OpenIoT Project, which is co-funded by the European Commission under the Seventh Framework Program, FP7-ICT-2011-7-287305-OpenIoT. The Author(s) acknowledge help and contributions from The Australian
National University.

\ifCLASSOPTIONcaptionsoff
  \newpage
\fi



%\bibliographystyle{IEEEtran}
%\begin{thebibliography}{1}
%\end{thebibliography}

%\vspace{-4pt}

\def\IEEEbibitemsep{0pt plus .5pt}
\bibliography{Bibliography}
\bibliographystyle{IEEEtran}

\vspace{-20pt}

\begin{IEEEbiography}[{\includegraphics[width=1in,height=1.25in,clip,keepaspectratio]{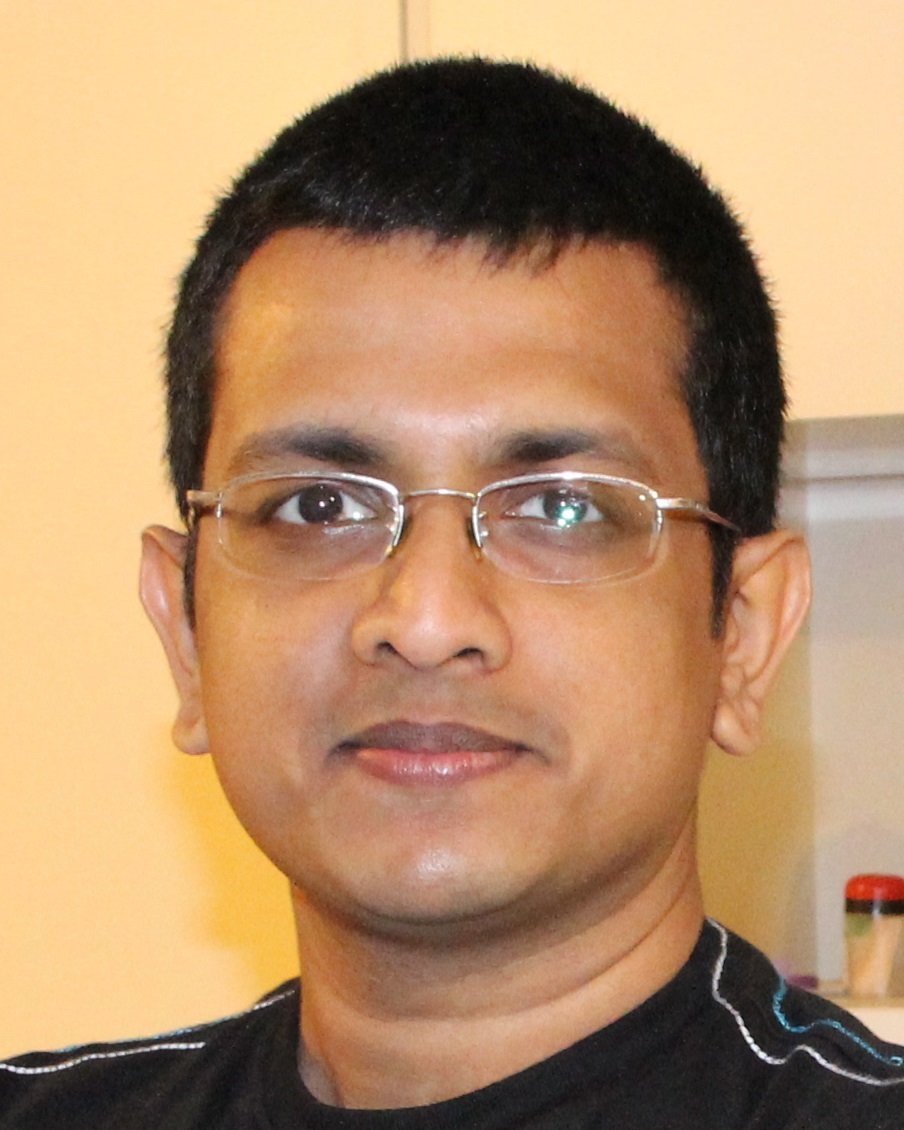}}]{Charith Perera}
received his BSc (Hons) in Computer Science in 2009 from Staffordshire University, Stoke-on-Trent, United Kingdom and MBA in Business Administration in 2012 from University of Wales, Cardiff, United Kingdom. He is currently pursing his PhD in Computer Science at The Australian National University, Canberra. He is also working at Information Engineering Laboratory, ICT Centre, CSIRO. His research interests include Internet of Things, pervasive and ubiquitous computing with a focus on sensor networks, and context aware computing. He is a member of the Association for Computing Machinery (ACM) and the Institute of Electrical and Electronics Engineers (IEEE).

\end{IEEEbiography}

\vspace{-30pt}

\begin{IEEEbiography}[{\includegraphics[width=1in,height=1.25in,clip,keepaspectratio]{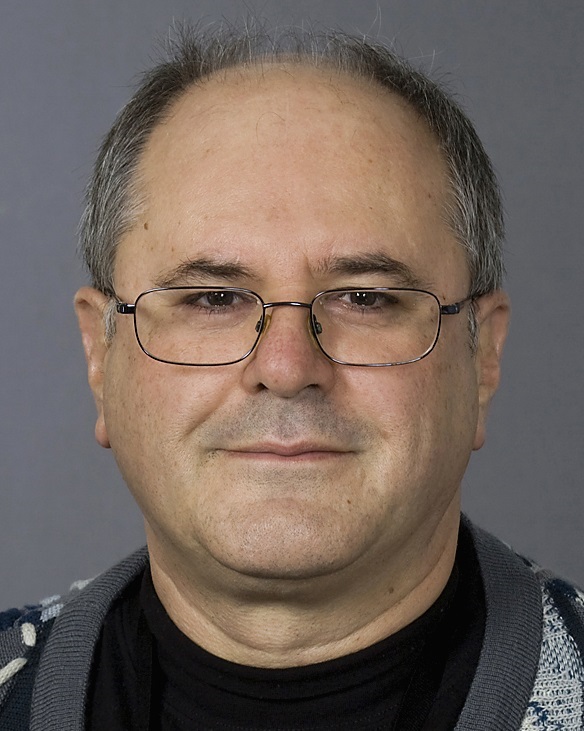}}]{Arkady Zaslavsky}
is the Science Leader of the Semantic Data Management science area at Information Engineering Laboratory, ICT Centre, CSIRO. He is also holding positions of Adjunct Professor at ANU, Research Professor at LTU and Adjunct Professor at UNSW. He is currently involved and is leading a number of European and national research projects.  Arkady received MSc in Applied Mathematics majoring in Computer Science from Tbilisi State University (Georgia, USSR) in 1976 and PhD in Computer Science from the Moscow Institute for Control Sciences (IPU-IAT), USSR Academy of Sciences in 1987.  Arkady Zaslavsky has published more than 300 research publications throughout his professional career. Arkady Zaslavsky is a Senior Member of ACM, a member of IEEE Computer and Communication Societies.
\end{IEEEbiography}

\vspace{-20pt}

\begin{IEEEbiography}[{\includegraphics[width=1in,height=1.25in,clip,keepaspectratio]{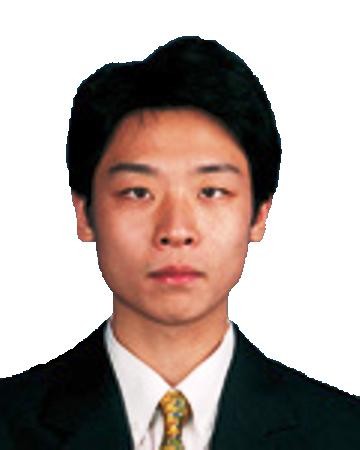}}]{Chi Harold Liu}
Chi Harold Liu is a staff researcher at IBM Research – China. He holds a Ph.D. degree from Imperial College, U.K., and a B.Eng. degree from Tsinghua University, China. His current research interests include the Internet-of-Things, big data analytics, mobile computing, and wireless ad-hoc, sensor and mesh networks. He receives the Distinguished Young Scholar Award in 2013, IBM First Plateau Invention Achievement Award in 2012, IBM First Patent Application Award in 2011, and interviewed by EEWeb.com as the Featured Engineer in 2011. He published widely in major conferences and journals, and owned 10 EU/US/China patents. He has served as the General Chair of international workshops with IEEE SECON'13, IEEE WCNC'12 and ACM UbiComp'11.

\end{IEEEbiography}

\vspace{-25pt}

\begin{IEEEbiography}[{\includegraphics[width=1in,height=1.25in,clip,keepaspectratio]{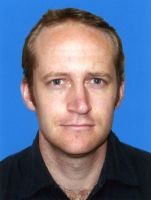}}]{Michael Compton}
is a research scientist in the Information Engineering Laboratory of CSIRO's ICT Centre. Since joining the ICT Centre in 2006 he has worked in the Information Security and Privacy team in the pHealth and Sensor Networks themes, in the Data Services for Sensor Networks project in the Sensor Networks theme, and now works in the Semantic Frameworks for Hydrological Sensor Webs project in the IWIS theme. His research interest is in using logic and specification to model and reason about systems and algorithms and as parts of systems that execute declarative descriptions of required processing. Currently he works with Semantic Web technologies for security, sensors and data integration. He received BIT (Hons) from The Australian National University in 2000 and PhD from University of Cambridge in 2007.
\end{IEEEbiography}

\vspace{-15pt}

\begin{IEEEbiography}[{\includegraphics[width=1in,height=1.25in,clip,keepaspectratio]{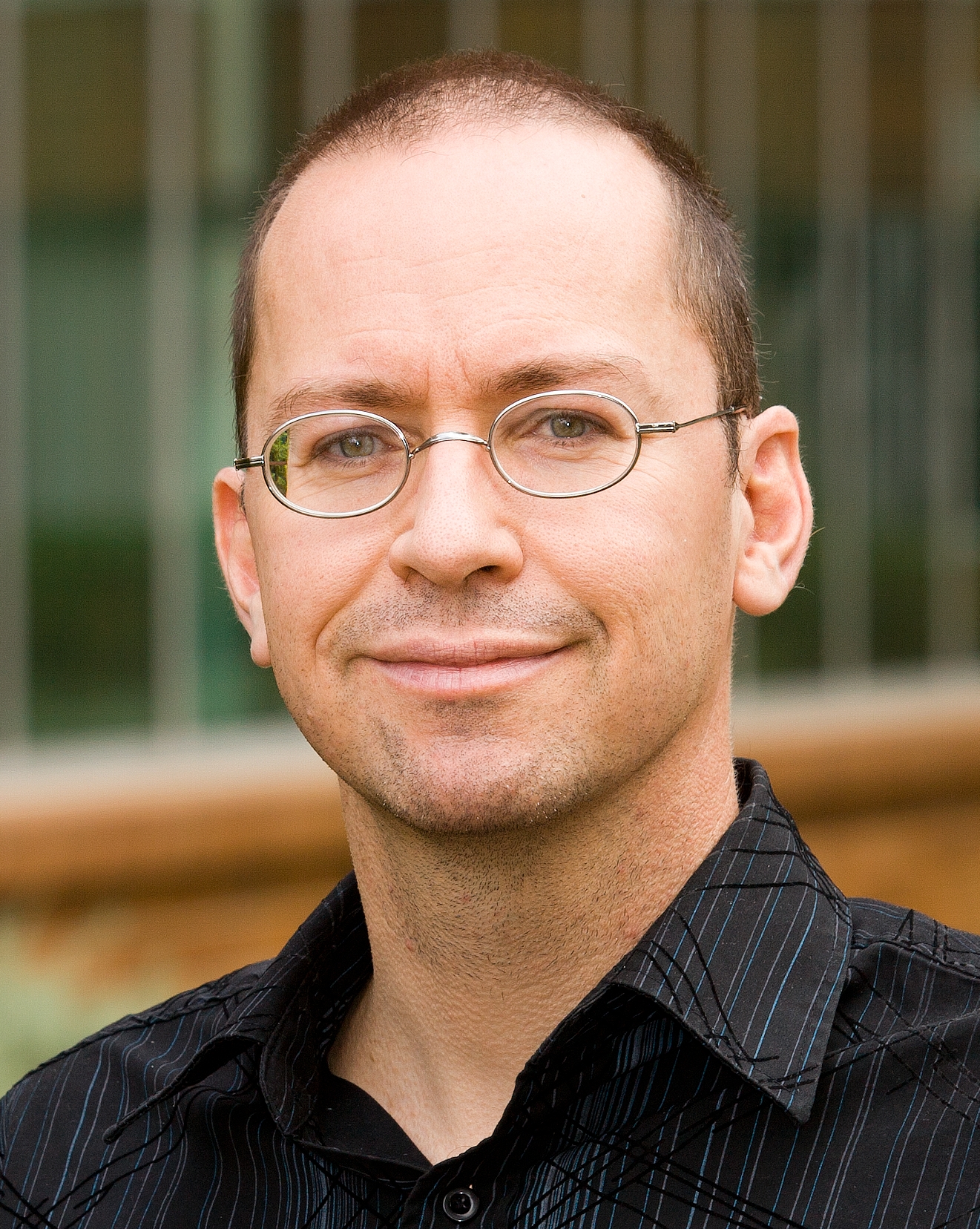}}]{Peter Christen}
is an Associate Professor in the Research School of Computer Science
at the Australian National University. He received his Diploma in Computer Science Engineering from ETH Z\"urich in 1995 and his PhD in Computer Science from the University of Basel in 1999 (both in Switzerland). His research interests are in data mining and data matching (entity resolution). He is especially interested in the development of scalable and real-time algorithms for data matching, and privacy and confidentiality aspects of data matching and data mining. He has published over 80 papers in these areas, including in 2012 the book `Data Matching' (by Springer), and he is the principle developer of the \emph{Febrl} (Freely Extensible Biomedical Record Linkage) open source data cleaning, deduplication and record linkage system.
\end{IEEEbiography}

\vspace{-350pt}

\begin{IEEEbiography}[{\includegraphics[width=1in,height=1.25in,clip,keepaspectratio]{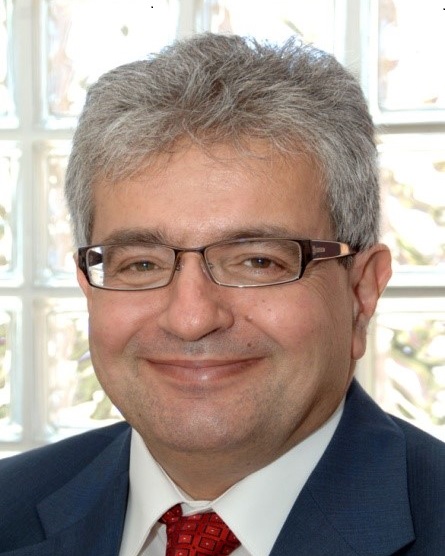}}]{Dimitrios Georgakopoulos}
is a Research Director at the CSIRO ICT Centre where he heads the Information Engineering Laboratory that is based in Canberra and Sydney.  Dimitrios is also an Adjunct Professor at the Australian National University. Before coming to CSIRO in October 2008, Dimitrios held research and management positions in several industrial laboratories in the US. From 2000 to 2008, he was a Senior Scientist with Telcordia, where he helped found Telcordia’s Research Centers in Austin, Texas, and Poznan, Poland. From 1997 to 2000, Dimitrios was a Technical Manager in the Information Technology organization of Microelectronics and Computer Corporation (MCC), and the Chief Architect of MCC’s Collaboration Management Infrastructure (CMI) consortial project. Dimitrios has received a GTE (Verizon) Excellence Award, two IEEE Computer Society Outstanding Paper Awards, and was nominated for the Computerworld Smithsonian Award in Science. He has published more than  hundred journal and conference papers.

\end{IEEEbiography}

%\begin{biography}[{\includegraphics[width=1in,height=1.25in,clip,keepaspectratio]{mshell}}]{Michael Shell}


%\vfill


\end{document}